\documentclass[aps,pre,showpacs,showkeys]{revtex4}

\usepackage{graphicx}                   
\usepackage{amsmath,amsfonts,amssymb}   

\newcommand{\bec}{\begin{center}}
\newcommand{\eec}{\end{center}}
\newcommand{\beq}{\begin{equation}}
\newcommand{\eeq}{\end{equation}}
\newcommand{\bea}{\begin{eqnarray}}
\newcommand{\eea}{\end{eqnarray}}
\newcommand{\half}{\frac{1}{2}}
\newcommand{\bfig}{\begin{figure}[h]}
\newcommand{\efig}{\end{figure}}
\newcommand{\bei}{\begin{itemize}}
\newcommand{\eei}{\end{itemize}}
\newcommand{\ben}{\begin{enumerate}}
\newcommand{\een}{\end{enumerate}}
\newcommand{\ba}{\begin{align}}
\newcommand{\ea}{\end{align}}
\newcommand{\eref}[1]{(\ref{#1})}

\newcommand{\RR}{\mathbb{R}}  
\newcommand{\kB}{k_B}         
\newcommand{\qmark}[1]{``#1''} 
\newcommand{\fig}[1]{Fig.~\ref{#1}} 

\newcommand{\Dim}{D}               
\newcommand{\Nvvec}{b}             
\newcommand{\Oherm}{K}             
\newcommand{\Nherm}{n}             
\newcommand{\vecd}[1]{\mathbf{#1}} 
\newcommand{\vecK}[1]{\underline{#1}} 
\newcommand{\Rb}{x}                
\newcommand{\Vb}{v}                
\newcommand{\Rd}{\vecd{\Rb}}       
\newcommand{\Vd}{\vecd{\Vb}}       
\newcommand{\ti}{t}                
\newcommand{\Dt}{\Delta\ti}        

\newcommand{\He}[2]{\bar{\mathcal{H}}_{#1}^{(#2)}} 
\newcommand{\HeHomo}[2]{\mathcal{H}_{#1}^{(#2)}} 
\newcommand{\Mf}[2]{M_{#1}^{(#2)}} 
\newcommand{\CHe}[2]{C_{#1}^{(#2)}} 
\newcommand{\CHef}[2]{F_{#1}^{(#2)}} 

\newcommand{\we}{\omega}           
\newcommand{\weD}{\we}             
\newcommand{\wi}{w}                
\newcommand{\wiMat}{W}             
\newcommand{\vT}{\Vb_T}            

\newcommand{\DiA}{\alpha}          
\newcommand{\DiB}{\beta}           
\newcommand{\DiC}{\gamma}          
\newcommand{\DiD}{\delta}          
\newcommand{\QiA}{i}               
\newcommand{\QiB}{j}               
\newcommand{\KiA}{l}               
\newcommand{\KiB}{m}               
\newcommand{\HiB}{k}               

\newcommand{\fB}{f}                
\newcommand{\fBB}[3]{\fB(#1,#2;#3)} 
\newcommand{\fBwe}{g}              
\newcommand{\fBweB}[3]{\fBwe_{#1}(#2;#3)} 
\newcommand{\fBn}{\rho}             
\newcommand{\fBJ}{J}                
\newcommand{\fBu}{u}                
\newcommand{\fBP}{P}                
\newcommand{\fBQ}{Q}                

\newcommand{\CopB}{\hat{C}}        
\newcommand{\Cop}[1]{\CopB[#1]}    
\newcommand{\CopC}[1]{\CopB\circ #1}  
\newcommand{\LCopB}{\hat{L}}       

\newcommand{\tauinv}{1/\tau}       
\newcommand{\tauinvDt}{\Dt/\tau}   

\newcommand{\gFP}{\gamma}          

\newcommand{\EF}{{E}}              
\newcommand{\aEF}{a^\EF}           
\newcommand{\uEF}{\fBu^\EF}        
\newcommand{\aEFd}{\vecd{a}^\EF}   

\newcommand{\ColMat}{\bar{C}}      

\newcommand{\Hnorm}{N}             
\newcommand{\IdMat}{I}             

\newcommand{\CEp}{\epsilon}        
\newcommand{\Dtheo}{D_0}           
\newcommand{\Dsim}{D}              

\newcommand{\dca}{\partial_{\alpha}}
\newcommand{\dcb}{\partial_{\beta}}
\newcommand{\dct}{\partial_{t}}
\newcommand{\da}{\partial_{\alpha}^{(1)}}
\newcommand{\db}{\partial_{\beta}^{(1)}}

\newcommand{\dt}{\partial_{t}^{(1)}}
\newcommand{\ddt}{\partial_{t}^{(2)}}
\newcommand{\ia}{_{i\alpha}}
\newcommand{\ib}{_{i\beta}}

\newcommand{\so}{^{(0)}}
\newcommand{\si}{^{(1)}}
\newcommand{\sii}{^{(2)}}
\newcommand{\st}{^{*}}

\begin{document}

\title{Solving the Fokker-Planck kinetic equation on a lattice}

\author{Daniele Moroni$^1$, Benjamin Rotenberg$^2$, 
Jean-Pierre Hansen$^1$, Sauro Succi$^{3}$, and Simone Melchionna$^4$}
\vskip 0.2cm

\affiliation{ 
$^{1}$ Department of Chemistry, University of Cambridge, \\
Lensfield Road, Cambridge CB2 1EW, United Kingdom \\
$^{2}$ Laboratoire Liquides Ioniques et Interfaces Charg\'{e}es, UMR 7612,
 Universit\'{e} Pierre et Marie Curie, 4 pl. Jussieu 75252 Paris Cedex 05 \\
$^{3}$ Istituto per le Applicazioni del Calcolo ``A. Picone'', \\
 CNR, Viale del Policlinico 137, 00161, Rome, Italy\\
$^{4}$ INFM-SOFT, Department of Physics, University of Rome ``La Sapienza'', \\
 P.le A. Moro 2, 00185 Rome, Italy}

\begin{abstract}
We propose a discrete lattice version of the Fokker-Planck kinetic equation
along lines similar to the Lattice-Boltzmann scheme. Our work extends an earlier
one-dimensional formulation to arbitrary spatial dimension $\Dim$. A generalized
Hermite-Gauss procedure is used to construct a discretized kinetic equation and
a Chapman-Enskog expansion is applied to adapt the scheme so as to correctly
reproduce the macroscopic continuum equations. The stability of the algorithm
with respect to the finite time-step $\Dt$ is characterized by the eigenvalues
of the collision matrix. A heuristic second-order algorithm in $\Dt$ is applied
to investigate the time evolution of the distribution function of simple model
systems, and compared to known analytical solutions. Preliminary investigations
of sedimenting Brownian particles subjected to an orthogonal centrifugal force
illustrate the numerical efficiency of the Lattice-Fokker-Planck algorithm to
simulate non-trivial situations. Interactions between Brownian particles may be
accounted for by adding a standard BGK collision operator to the discretized
Fokker-Planck kernel.
\end{abstract}
\vskip 0.2cm


\maketitle

\section{Introduction}

Kinetic equations are well established mathematical models for investigating the
out of equilibrium behaviour of fluids, and their relaxation towards
thermodynamic equilibrium, at a molecular or coarse-grained, mesoscopic
level~\cite{ResiboisLeener}.  They govern the time evolution of the single
particle distribution function $\fBB{\Rd}{\Vd}{\ti}$ in the 2$\times
D$-dimensional space of position $\Rd$ and velocity $\Vd$. This evolution is
expressed in terms of \qmark{free flow}, under the action of an external or
self-consistent force field, and of the action of a \qmark{collision} operator
$\Cop{\fB}$, which accounts for the interactions between particles, or their
coupling to a continuous medium. The exact form of the operator $\CopB$ involves
a hierarchy of equations for the higher order distribution functions (the BBGKY
hierarchy~\cite{HansenMcDonald}), so that a closed equation for $\fB$ cannot be
obtained. Depending on the physical problem at hand, approximate closures have
been devised which lead to various standard kinetic equations.

Thus, if particle interactions are only considered at a mean-field level,
through a self-consistent force field, $\CopB\equiv 0$ and the standard Vlasov
equation of plasma physics results~\cite{MontgomeryTidman}. In dilute gases of
molecules interacting through short-range forces, one may make the assumption of
strictly binary, uncorrelated collisions, which leads to the non-linear
Boltzmann collision operator involving the molecular scattering
cross-section~\cite{ResiboisLeener}. The Boltzmann equation has been widely used
for a systematic investigation of transport phenomena in
gases~\cite{Cercignani}, while its generalization by Enskog, which accounts for
static correlations, allows such calculations to be extended to dense
fluids~\cite{ResiboisLeener}.

Following the idea that the molecular details included in the Boltzmann and
Enskog collision operators are not likely to have a strong influence on the
experimentally measured macroscopic properties of fluids, Bhatnagar, Gross and
Krook (BGK) proposed a highly simplified, phenomenological version of the
collision operator describing the relaxation of the distribution function
towards local Maxwellian equilibrium on a single time scale $\tau$.  The BGK
operator~\cite{BGK54} still conserves mass, momentum and kinetic energy, and by
properly adjusting the relaxation time $\tau$, it goes beyond the strictly
binary collision assumption of the Boltzmann equation and hence allows its
phenomenological extension to dense fluids~\cite{Succi}. The combination of the
BGK kernel with discretized lattice versions of kinetic equations, globally
referred to as the \emph{Lattice Boltzmann} (LB)
method~\cite{BSV92,WolfGladrow,Succi} has proved to be a powerful tool for the
study of laminar or turbulent fluid flow and transport. The assumption that
fluid particles can be restricted to have only a small, fixed number of
velocities $\Vd$ reduces the computationtal problem considerably compared to
corresponding finite element schemes.  The numerical parameters of the
Lattice-BGK model can be adjusted to reproduce the correct Navier-Stokes
behaviour in the small velocity (or Mach number) limit. Over the last decade the
LB method has increased in popularity as successful applications have been
repeatedly reported in numerical simulations of large scale hydrodynamic
flows~\cite{CKOSSY03}, complex fluids under shear and in porous
media~\cite{HCVC05}, ion transport in nanochannels~\cite{MS05}, and colloidal
suspensions~\cite{LV01,SAPDC05}.

The latter systems generally involve a considerable separation in size and time
scales as epitomized by the classic concept of Brownian motion. This is
generally the case of two-component systems involving a molecular-scale solvent
and larger, heavier solutes. The natural kinetic theory framework to handle such
highly asymmetric situations is the Fokker-Planck (or Kramers)
equation~\cite{Risken}, which adopts an effective, one-component description of
the solute, whereby the solvent and the boundaries are modelled implicitly, as
sources of friction and random forces.  The collision operator $\Cop{\fB}$ may
then be constructed as the sum of a Fokker-Planck (FP) operator, which accounts
for the coupling between the solute and the (continuous) solvent and between the
solute and the confining surfaces, and of a BGK operator to model solute/solute
interactions. The corresponding lattice Fokker-Planck (LFP) equation was
recently put forward by some of us~\cite{MSH05}, and applied to a simple
one-dimensional problem of electrical conduction.

The main objective of the present paper is to extend the LFP formulation to the
$\Dim$-dimensional case, and to develop an efficient and stable numerical scheme
for its solution. This should provide an operational tool to tackle
non-equilibrium problems in the field of dispersions and complex fluids
involving multiple length and time scales.

The paper is organized as follows.  The lattice discretization of the FP
equation is carried out in sec.~\ref{sec:LFP}.  Using a truncated expansion of
the distribution function $\fBB{\Rd}{\Vd}{\ti}$ in generalized Hermite
polynomials, the collision operator $\Cop{\fB}$ is expressed in terms of the
moments of $\fB$, which can be computed by appropriate quadratures. This
completely defines the LFP numerical solution scheme.  In sec.~\ref{sec:colli}
we address the stability of such a scheme. Since the evolution of the
discretized distribution functions can be rewritten as a linear iteration,
studying the stability amounts to analysing the spectrum of the transformation. 
A standard Chapman-Enskog expansion of the LFP equation is carried out in
sec.~\ref{sec:ChapEns} to ascertain the reproducibility of the continuous
macroscopic equations.  The practical implementation of a second order algorithm
in the discrete time step $\Dt$ is proposed in sec.~\ref{sec:corrected}, while
numerical results are presented in sec.~\ref{sec:numerics}.  Concluding remarks
are contained in sec~\ref{sec:conclusion}, and mathematical issues are detailed
in the appendices.

\section{The Lattice Fokker-Planck Equation}
\label{sec:LFP}\label{sec:quadra}\label{sec:hermite}

Since the implementation of the BGK collision operator in the LB method is amply
documented in the literature~\cite{WolfGladrow,Succi}, we restrict the following
to the Fokker-Planck operator. The standard FP kinetic equation in $\Dim$
dimensions reads~\cite{Risken}:
\beq\label{eq:FPE1}
(\partial_\ti + \Vb_\DiA\partial_{\DiA} + \aEF_\DiA \partial_{\Vb_\DiA}) \fB = 
\CopB^{FP}[\fB] \doteq \gFP\partial_{\Vb_\DiA} 
 (\Vb_\DiA + \vT^2\partial_{\Vb_\DiA}) \fB
\eeq
where $\Rb_\DiA$ and $\Vb_\DiA$ are the cartesian components of the
$\Dim$-dimensional position and velocity vectors, $\partial_{\DiA}$ and
$\partial_{\Vb_\DiA}$ the corresponding gradient operators, and $\aEF_\DiA$ are
the components of the acceleration due to an external force field $m \aEFd$
acting on the solute particles of mass $m$.  Here and in the following greek
indices run from 1 to $\Dim$ and we adopt Einstein's summation convention over
repeated indices.  The left-hand side is a conventional streaming operator,
while the right-hand side is a Fokker-Planck operator with constant friction
coefficient $\gFP$ and thermal velocity $\vT^2=\kB T/m$, where $\kB$ is the
Boltzmann constant and $T$ the temperature of the system.

\begin{figure}
\begin{center}
\includegraphics[]{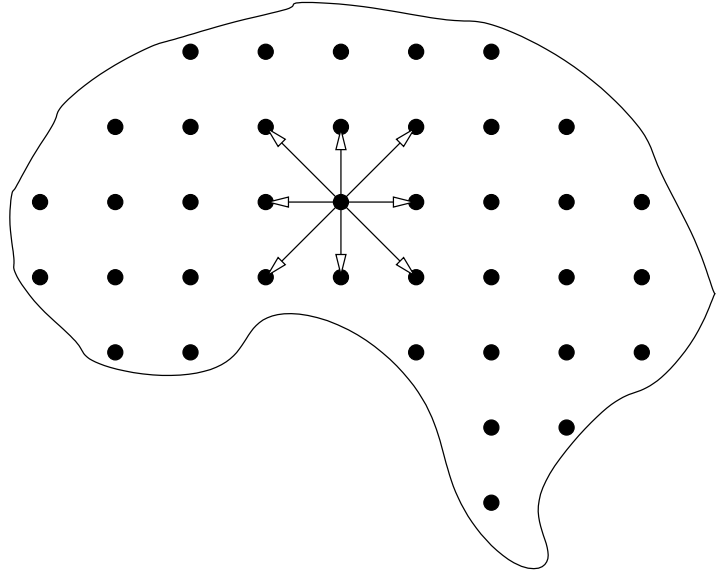}
\caption{The discretization of Lattice Boltzmann methods. A lattice in $\Rd$
space is chosen and the velocity $\Vd$ is restricted to a finite set $\Vd_\QiA$,
$\QiA=1\ldots\Nvvec$ of lattice vectors. The choice of the lattice and the
$\Vd_\QiA$ is not arbitrary, but is constructed to allow the calculation of the
moments of $\fB$ by quadratures. In the figure a two-dimensional cubic lattice
is depicted with 9 velocity vectors, usually called the D2Q9 model.}
\label{fig:lattice}
\end{center}
\end{figure}

In the Lattice-Boltzmann method the continuous velocity $\Vd$ is replaced by a
finite set of discrete velocities $\Vd_\QiA$, $\QiA=1\ldots\Nvvec$ which are
vectors on a lattice. Accordingly, the distribution function
$\fBB{\Rd}{\Vd}{\ti}$ is replaced by $\Nvvec$ functions
$\fBweB{\QiA}{\Rd}{\ti}\propto\fBB{\Rd}{\Vd_\QiA}{\ti}$, and Eq.~\eref{eq:FPE1}
by $\Nvvec$ equations for each of the $\fBwe_\QiA$. In these equations the
$\Vd_\QiA$ are no longer variables, but fixed parameters. The positions $\Rd$
are discrete points on the lattice, whose size and boundaries are modelled on
the geometry of the physical problem. Completing the discretization, time $\ti$
is considered to vary in multiples of a discrete step $\Dt$.  This passage from
a continuous to a discrete world is schematically pictured in
\fig{fig:lattice}.
In order to derive the discrete equations,
we define more conveniently the external force operator
\bea
\CopB^{EXT}[\fB] &=& -\aEF_\DiA\partial_{\Vb_\DiA} \fB
\eea
and rewrite equation~\eref{eq:FPE1} as
\beq\label{eq:FPE2}
(\partial_\ti + \Vb_\DiA\partial_{\DiA}) \fB = \Cop{\fB} 
\eeq 
with $\Cop{\fB} = \CopB^{FP}[\fB]+\CopB^{EXT}[\fB]$.  On the left-hand side we
now have a free-particle streaming operator, which can be easily discretized, as
we will show in the next section. The non-trivial task is to find the correct
lattice collision operator $\LCopB$ corresponding to the continuous operator
$\CopB$. In the BGK case, a systematic procedure has been devised
in~\cite{SH98,HL97b} based on Gauss-Hermite quadratures. The procedure is
inspired by the pioneering ideas by Grad~\cite{Grad49b} to solve the Boltzmann
equation using the so-called 13-moment system, and has become a useful tool in
discrete models of the Boltzmann equation~\cite{AK02,AKOe03}.  It relies on the
fact that products of a gaussian with Hermite polynomials are eigenfunctions of
the BGK operator. We prove in the following that the Gauss-Hermite strategy also
allows to discretize the Fokker-Planck kinetic equation because BGK and FP
operators share the same set of eigenfunctions.  Indeed they are just different
limits of a more general integral operator, devised by Skinner and
Wolynes~\cite{SW80}.  The additional force term present in $\Cop{\fB}$ is not
diagonal in this basis set, but it has nevertheless a computationally convenient
form.  Although the methodology is not new, its application in the FP context is
the novelty of the present work. In order to provide a comprehensive and
self-contained treatment we give full mathematical details in the following.

The discretization is best carried out by expanding the continuous distribution
function $\fBB{\Rd}{\Vd}{\ti}$ over a basis set of $\Dim$-dimensional Hermite
polynomial tensors $\HeHomo{\vecK{\DiA}}{\KiA}(\Vd)$ (see
appendix~\ref{sec:hermitepoly}), according to
\beq\label{eq:HermiExpa0}
\fBB{\Rd}{\Vd}{\ti}= \weD(\Vd)
 \sum_{\KiA=0}^\infty \frac{1}{\vT^{2\KiA}\KiA!} 
 \CHef{\vecK{\DiA}}{\KiA}(\Rd,\ti) \HeHomo{\vecK{\DiA}}{\KiA}(\Vd)
\eeq
where the subscript $\vecK{\DiA}$ is an abbrevation for $\DiA_1\ldots\DiA_\KiA$,
the product denotes contraction on all $\KiA$ indices, and
\beq\label{eq:gaussweight}
\weD(\Vd)=\frac{e^{-\Vb^2/(2\vT^2)}}{(2\pi\vT^2)^{\Dim/2}}
\eeq
is a gaussian weight function, with $\Vb^2=\Vd\cdot\Vd$.
The expansion coefficients are given by
\beq\label{eq:HermiCoeff}
\CHef{\vecK{\DiA}}{\KiA}(\Rd,\ti)= \int d\Vd \fBB{\Rd}{\Vd}{\ti}
 \HeHomo{\vecK{\DiA}}{\KiA}(\Vd)
\eeq
where the velocity integrals are always taken over $\RR^\Dim$.
Since $\HeHomo{\vecK{\DiA}}{\KiA}(\Vd)$ involve polynomials of order $\KiA$,
the $\CHef{\vecK{\DiA}}{\KiA}$ are linear combinations of the moments of $\fB$
\beq\label{eq:fmom}
\Mf{\vecK{\DiA}}{\KiB}(\Rd,\ti)=
 \int d\Vd \fBB{\Rd}{\Vd}{\ti} \Vb_{\DiA_1}\ldots\Vb_{\DiA_\KiB}
\eeq
with $\KiB\leq\KiA$.

Inserting the expansion in the kinetic equation~\eref{eq:FPE2}, we could project
the equation on the basis set and derive a hierarchy of differential equations
for the coefficients. However, that would simply transform the problem into
another of equivalent complexity.  Here we aim instead at a different
approach. By means of the Hermite expansion we can express the right-hand side
of~\eref{eq:FPE2} as a function of the moments of $\fB$. Using a
lattice-discretized distribution function we can compute these moments by a
suitable quadrature. The expansion involves an infinite number of moments and
naturally we have to truncate it at a certain order $\Oherm$. We hence
\emph{assume} that
\beq
\CHef{\vecK{\DiA}}{\KiA}(\Rd,\ti)=0 \quad\text{if $\KiA>\Oherm$}
\eeq
and rewrite $\fB$ as a distribution function that lies entirely
in the subspace of Hermite polynomials up to order $\Oherm$
\beq\label{eq:HermiExpa3}
\fBB{\Rd}{\Vd}{\ti}=\weD(\Vd)
 \sum_{\KiA=0}^\Oherm \frac{1}{\vT^{2\KiA}\KiA!} 
 \CHef{\vecK{\DiA}}{\KiA}(\Rd,\ti) \HeHomo{\vecK{\DiA}}{\KiA}(\Vd)
\eeq
This assumption is expected to be valid at least for situations close to
equilibrium.  Using the properties of $\CopB^{FP}[\fB]$ and $\CopB^{EXT}[\fB]$
detailed in appendix~\ref{sec:eigenhermite}, we also find that the outcome of
$\Cop{\fB}=\CopB^{FP}[\fB]+\CopB^{EXT}[\fB]$ lies entirely in this subspace.

The key idea of the Lattice Boltzmann method is that in order to compute the
velocity integrals in Eq.~\eref{eq:fmom}, only a finite set of velocities is
needed. Specifically, we want to compute $\Dim$-dimensional integrals as
discretized sums over a fixed set of points. We assume there exists a set of
vectors $\Vd_\QiA\in\RR^\Dim$, and a set of real numbers $\wi_\QiA$, with
$\QiA=1\ldots\Nvvec$, such that if $p(\Vd)$ is a polynomial of degree not
greater than $2\Oherm$, the following formula is valid
\beq\label{eq:quadra}
\int d\Vd \weD(\Vd)p(\Vd) = \sum_{\QiA=1}^{\Nvvec} \wi_\QiA p(\Vd_\QiA)
\eeq
The above equation is then called a quadrature of degree $2\Oherm$, and the
$\Vd_\QiA$, $\wi_\QiA$, the nodes and weights of the quadrature. 

Because of the truncated expansion Eq.~\eref{eq:HermiExpa3}, $\fB/\weD$ is a
polynomial of order $\Oherm$ at most.  Since we requested a quadrature of degree
$2\Oherm$ we can now compute the moments of $\fB$ up to order $\Oherm$,
according to
\bea\label{eq:Lfmom}
&& \int d\Vd \fBB{\Rd}{\Vd}{\ti} \Vb_{\DiA_1}\ldots\Vb_{\DiA_\KiB}
=\int d\Vd \frac{\weD(\Vd)}{\weD(\Vd)}
 \fBB{\Rd}{\Vd}{\ti}\Vb_{\DiA_1}\ldots\Vb_{\DiA_\KiB}= \\
&&\quad =\sum_{\QiA=1}^{\Nvvec} \wi_\QiA 
 \frac{\fBB{\Rd}{\Vd_\QiA}{\ti}}{\weD(\Vd_\QiA)}
 \Vb_{\QiA\DiA_1}\ldots\Vb_{\QiA\DiA_\KiB}
\doteq \sum_{\QiA=1}^{\Nvvec} \fBwe_\QiA\Vb_{\QiA\DiA_1}
 \ldots\Vb_{\QiA\DiA_\KiB} \nonumber
\eea
where we defined $\fBweB{\QiA}{\Rd}{\ti}=\wi_\QiA
\fBB{\Rd}{\Vd_\QiA}{\ti}/\weD(\Vd_\QiA)$ and the formula is valid for
$m\leq\Oherm$.

The choice of $\Oherm$ is dictated by the application. In practice one is not
interested in having $\fB$ itself, but rather compute its moments, which
correspond to macroscopic observables. Momentum and energy equations involve
moments up to second and third order respectively.  Consequently it is necessary
to require $\Oherm\geq 2$ or $\Oherm\geq 3$. The lower order moments are
labelled by conventional names and the quadratures read
\begin{subequations}
\label{eq:LfmomE}
\bea
\fBn \doteq \Mf{}{0}
 = \int d\Vd \fBB{\Rd}{\Vd}{\ti} 
 &=& \sum_{\QiA=1}^{\Nvvec} \fBwe_\QiA \label{eq:LfmomE0} \\
\fBJ_\DiA\doteq\fBn\fBu_\DiA \doteq \Mf{\DiA}{1}
 = \int d\Vd \fBB{\Rd}{\Vd}{\ti} \Vb_\DiA 
 &=& \sum_{\QiA=1}^{\Nvvec} \fBwe_\QiA \Vb_{\QiA\DiA} \\
\fBP_{\DiA\DiB} \doteq \Mf{\DiA\DiB}{2}
 = \int d\Vd \fBB{\Rd}{\Vd}{\ti} \Vb_\DiA\Vb_\DiB 
 &=& \sum_{\QiA=1}^{\Nvvec} \fBwe_\QiA \Vb_{\QiA\DiA}\Vb_{\QiA\DiB} 
\eea
and if $\Oherm\geq 3$ we can also compute exactly
\bea
\fBQ_{\DiA\DiB\DiC} \doteq \Mf{\DiA\DiB\DiC}{3}
 = \int d\Vd \fBB{\Rd}{\Vd}{\ti} \Vb_\DiA\Vb_\DiB\Vb_\DiC 
 &=& \sum_{\QiA=1}^{\Nvvec} \fBwe_\QiA 
  \Vb_{\QiA\DiA}\Vb_{\QiA\DiB}\Vb_{\QiA\DiC}
\eea
\end{subequations}

Finding the optimal set $\{\Vd_\QiA$, $\wi_\QiA\}$ in terms of a minimum number
of nodes for a given degree of accuracy, is in general an unsolved
problem~\cite{Stroud}. However, as far as the solutions of kinetic equations are
concerned, it is also important that the $\Vd_\QiA$ be vectors of a regular
lattice in $\Rd$ space, as shown in \fig{fig:lattice}.  Then for the cases of
physical interest ($\Dim=1,2,3$ and $\Oherm=2,3$) a number of possibilities
exist with different $\Nvvec$'s.  The thermal velocity $\vT$ can also become a
free parameter to adjust the quadratures.  Some resulting models that are used
in practice can be found for example in~\cite{Succi}.

We now have the prerequisites to create a computational scheme for the
Fokker-Planck kinetic equation. Consider the distribution function $\fB$
evaluated at discrete lattice points $\Rd$ and at a finite set of $\Nvvec$
velocities $\Vd_\QiA$ that are also lattice vectors. Multiplying both sides of
Eq.~\eref{eq:FPE2} by $\wi_\QiA/\weD(\Vd_\QiA)$ and using the expansion
\eref{eq:HermiExpa3} not on $\fB$ but on the function $\Cop{\fB}$, we can write
\beq
\partial_\ti \fBwe_\QiA + \Vb_{\QiA\DiA}\partial_{\DiA} \fBwe_\QiA 
 = \wi_\QiA
 \sum_{\KiA=0}^\Oherm \frac{1}{\vT^{2\KiA}\KiA!} 
 \CHe{\vecK{\DiA}}{\KiA} \HeHomo{\vecK{\DiA}}{\KiA}(\Vd_\QiA) 
\eeq
where now
\beq\label{eq:CHedefi}
\CHe{\vecK{\DiA}}{\KiA}= 
 \int d\Vd \Cop{\fB} \HeHomo{\vecK{\DiA}}{\KiA}(\vecd{\Vd})
\eeq
Finite difference time discretization to first-order~\cite{HL97b} then leads to
\beq\label{eq:LB}
\fBweB{\QiA}{\Rd+\Vd_\QiA\Dt}{\ti+\Dt}-\fBweB{\QiA}{\Rd}{\ti} = \Dt\:\wi_\QiA
\sum_{\KiA=0}^\Oherm \frac{1}{\vT^{2\KiA}\KiA!} 
 \CHe{\vecK{\DiA}}{\KiA} \HeHomo{\vecK{\DiA}}{\KiA}(\Vd_\QiA) 
\eeq
which defines the lattice Fokker-Planck equation.  

The above expression must be supplemented with an operational expression for the
Hermite coefficients $\CHe{\vecK{\DiA}}{\KiA}$ of $\Cop{\fB}$.  Using the
results of appendix~\ref{sec:eigenhermite}, they can be expressed as functions
of the Hermite coefficient $\CHef{\vecK{\DiA}}{\KiA}$ of $\fB$. We can then
write
$\CHe{\vecK{\DiA}}{\KiA}=C_{\vecK{\DiA}}^{(\KiA),FP}+
 C_{\vecK{\DiA}}^{(\KiA),EXT}$, where
\begin{subequations}\label{eq:HcoeffC}
\bea
C_{\vecK{\DiA}}^{(\KiA),FP} &=& -\gFP\KiA\CHef{\vecK{\DiA}}{\KiA} 
 \label{eq:HcoeffC1} \\
C_{\vecK{\DiA}}^{(\KiA),EXT} &=& 
 \aEF_{\DiA_1}\CHef{\DiA_2\ldots\DiA_\KiA}{\KiA-1}+
 \ldots+
 \aEF_{\DiA_\KiA}\CHef{\DiA_1\ldots\DiA_{\KiA-1}}{\KiA-1} 
 \label{eq:HcoeffC2}
\eea
\end{subequations}
and $C_{\vecK{\DiA}}^{(0),EXT}=0$.  The $\CHef{\vecK{\DiA}}{\KiA}$ are related
in turn to the moments $\Mf{\vecK{\DiA}}{\KiA}$ of $\fB$ by the definition of
the Hermite polynomials.  For $\Oherm \leq 3$ they read
\begin{subequations}\label{eq:HcoeffF}
\bea
\CHef{}{0} &=& \fBn \\
\CHef{\DiA}{1} &=& \fBJ_\DiA \\
\CHef{\DiA\DiB}{2} &=& \fBP_{\DiA\DiB}-\vT^2\fBn\delta_{\DiA\DiB} \\
\CHef{\DiA\DiB\DiC}{3} &=& \fBQ_{\DiA\DiB\DiC}-
 \vT^2[\delta_{\DiA\DiB}\fBJ_\DiC+\delta_{\DiA\DiC}\fBJ_\DiB+
 \delta_{\DiB\DiC}\fBJ_\DiA]
\eea
\end{subequations}
where $\fBn,\fBJ_\DiA$, etc. are related to
the $\fBwe_\QiA$ via the quadratures Eqs.\eref{eq:LfmomE}.

Putting together Eqs.~\eref{eq:LB}, \eref{eq:HcoeffC}, \eref{eq:HcoeffF} and
\eref{eq:LfmomE}, we have then a complete numerical scheme to solve the
continuous equation~\eref{eq:FPE2}. More precisely, we have different schemes
according to the choice of the order $\Oherm$ in the Hermite expansion, which
allow corresponding exact computation of the moments of $\fB$ up to the same
order.  As an example we can write explicitly to second order ($\Oherm=2$)
\begin{subequations}\label{eq:LB2}
\beq\label{eq:LB2a}
\fBweB{\QiA}{\Rd+\Vd_\QiA\Dt}{\ti+\Dt}-\fBweB{\QiA}{\Rd}{\ti} 
 = \Dt\:\LCopB[\fBwe_\QiA]
\eeq
where the lattice collision operator $\LCopB$ reads
\beq\label{eq:LB2b}
\LCopB[\fBwe_\QiA] = \wi_\QiA\left\{
[-\gFP\fBJ_\DiA+\aEF_\DiA\fBn]\frac{\Vb_{\QiA\DiA}}{\vT^2}
+[-2\gFP(\fBP_{\DiA\DiB}-\vT^2\fBn\delta_{\DiA\DiB})
 +\aEF_\DiA\fBJ_\DiB+\aEF_\DiB\fBJ_\DiA]
\frac{\Vb_{\QiA\DiA}\Vb_{\QiA\DiB}-\vT^2\delta_{\DiA\DiB}}{2\vT^4}
\right\}
\eeq
\end{subequations}
Clearly, given the functions $\fBweB{\QiA}{\Rd}{\ti}$ at time $\ti$, one can
compute the moments $\fBn,\fBJ_\DiA,\fBP_{\DiA\DiB}$ and hence
$\LCopB[\fBwe_\QiA]$. The $\fBweB{\QiA}{\Rd}{\ti+\Dt}$ at time $\ti+\Dt$ are
then obtained using the left hand side of Eq.~\eref{eq:LB2a}. Note that in this
way we are not calculating the distribution function $\fB$ but rather
$\fBwe_\QiA=\wi_\QiA \fBB{\Rd}{\Vd_\QiA}{\ti}/\weD(\Vd_\QiA)$. However, the
quantities of interest to be sampled are the moments of $\fB$, corresponding to
hydrodynamic observables.  By construction the quadratures provide them
straightforwardly via Eqs.\eref{eq:LfmomE}.

The scheme derived here defines an algorithm for the numerical solution of
Eq.~\eref{eq:FPE2}. In the case $\Dim=1$, greek subscripts are no longer
necessary, expressions~\eref{eq:HcoeffC} reduce to
$\CHe{}{\KiA}=-\gFP\KiA\CHef{}{\KiA}+\aEF\KiA\CHef{}{\KiA-1}$, and
Eqs.~\eref{eq:HcoeffF} simplify as well. These expressions coincide with those
used in~\cite{MSH05} for a second order ($\Oherm=2$) scheme.  In~\cite{MSH05}
the discrete lattice equations are tested against the continuous equation only
numerically.  In this paper we intend to give a full analysis of the scheme
defined by Eq.~\eref{eq:LB2}.  Before giving details of the implementation, we
check in the following sections how reliable the present method is by addressing
its stability and the reproducibility of the continuous equation~\eref{eq:FPE2}.
The discussion of the computational algorithm is then postponed to
sec.~\ref{sec:corrected}.

\section{Stability analysis}\label{sec:colli}

In the following we show that Eq.\eref{eq:LB} can be recast in the linear form
\beq\label{eq:ColMat}
\fBwe_\QiA' = \sum_{\QiB} \ColMat_{\QiA\QiB}\fBwe_\QiB
\eeq
where
$\fBwe_\QiA'=\fBweB{\QiA}{\Rd+\Vd_\QiA\Dt}{\ti+\Dt}-\fBweB{\QiA}{\Rd}{\ti}$
defines a vector $\fBwe'$, $\fBwe_\QiB=\fBwe_\QiB(\Rd,\ti)$ a vector $\fBwe$,
and $\ColMat_{\QiA\QiB}$ the so-called collision matrix
$\ColMat$~\cite{HJ89,HSB89,Succi}.  For a given lattice geometry, $\ColMat$ is a
constant matrix which depends only on the operator parameters
$\gFP,\aEF_\DiA$. In particular we consider isothermal models, where $\vT$ is
fixed and we make the significant assumption that the external acceleration
field $\aEFd$ does not depend self-consistently on the distribution functions
$\fBwe_\QiA$.  The latter case will be examined in a subsequent publication.

The aim of this section is to check for which range of these parameters the
scheme embodied in Eq.~\eref{eq:ColMat} is stable, where stability means that
upon iterating the scheme, the distribution functions $\fBweB{\QiA}{\Rd}{\ti}$
stay finite at any value of $\Rd$ and $\ti$.  The task is substantially
facilitated by the fact that we do not have to first linearize the scheme, as in
usual Von Neumann stability analysis~\cite{WolfGladrow}.  By standard arguments
in Lattice Boltzmann theory~\cite{BSV92} the stability condition reads
\beq\label{eq:stable}
|1+\lambda(\ColMat)|<1
\eeq
where $\lambda(\ColMat)$ is any eigenvalue of $\ColMat$. We remark that
Eq.~\eref{eq:stable} is a very simple condition valid globally, independently of
the initial distributions $\fBweB{\QiA}{\Rd}{0}$ or the boundary geometry.  Such
a feature is an attractive consequence of the linearity of the scheme, while
much more complicated stability analyses which depend on the local
$\fBweB{\QiA}{\Rd}{0}$ are required in the full self-consistent LB
method~\cite{WMS97}.

Thanks to Eq.~\eref{eq:stable} the stability analysis reduces to the spectral
analysis of $\ColMat$.  We proceed now to first identify this matrix, and then
compute its spectrum using the results of the previous section.

As a starting point, we rewrite Eq.~\eref{eq:LB} as
\beq\label{eq:ColVec}
\fBwe_\QiA' / \Dt = \wi_\QiA
\sum_{\KiA=1}^{\Nherm} \frac{1}{\Hnorm_\KiA^2} 
 \CHe{}{\KiA} \HeHomo{}{\KiA}(\Vd_\QiA) 
\eeq
where temporarily in this section we set aside the tensorial notation and
enumerate all the terms of the sum (including the tensorial contraction) simply
from 1 to $\Nherm$. So the index $\KiA$ here represents a shorthand notation for
the previous set of indices $\KiA\vecK{\DiA}$.  Accordingly we redefine the
normalization factors as just $\Hnorm_\KiA^2$, since it is not necessary to know
their detailed form.  We wish then to express Eq.~\eref{eq:ColVec} in the matrix
form of Eq.~\eref{eq:ColMat} using the fact that the dependence on
$\fBwe_\QiA=\wi_\QiA/\weD(\Vd_i)\fB_i$ is inside $\CHe{}{\KiA} = \int d\Vd
\Cop{\fB}\HeHomo{}{\KiA}(\Vd)$.

The first step is then to use the quadratures to write 
(see appendix \ref{sec:hermitepoly})
\beq\label{eq:HeNorm}
\sum_\QiA \wi_\QiA \HeHomo{}{\KiA}(\Vd_\QiA) \HeHomo{}{\KiB}(\Vd_\QiA) 
 = \delta_{\KiA\KiB} \Hnorm_\KiA^2
\eeq
Defining the matrix $H_{\QiA\KiA}\doteq\HeHomo{}{\KiA}(\Vd_\QiA)$ (which
contains $\Nvvec$ rows times $\Nherm$ columns), Eq.~\eref{eq:HeNorm} is
rewritten in matrix form
\beq\label{eq:Hmat}
H^{T}\wiMat H = \Hnorm^2
\eeq
where $H^T$ is the transpose of $H$,
$\wiMat_{\QiA\QiB}\doteq\wi_\QiA\delta_{\QiA\QiB}$ is a $\Nvvec\times\Nvvec$
diagonal matrix, and $\Hnorm^2=\Hnorm_\KiA^2\delta_{\KiA\KiB}$ is a
$\Nherm\times\Nherm$ diagonal matrix.  Stated otherwise $H^T(\wiMat
H\Hnorm^{-2})=\IdMat$, i.e. $\wiMat H\Hnorm^{-2}$ is a right-inverse
$(H^T)^{-1,R}$ of $H^T$.

As a second step, we consider the operator $\CopB$, and we apply it to $\fB$
expanded in its Hermite representation
\beq
\CopC{\fB}=
 \CopC{}\left[\weD(\Vd)\sum_{\KiA=1}^{\Nherm} \frac{1}{\Hnorm_\KiA^2} 
 F^{(\KiA)} \HeHomo{}{\KiA}(\Vd)\right]=
\sum_{\KiA=1}^{\Nherm}\CopC{}\left[\weD(\Vd)
 \frac{\HeHomo{}{\KiA}(\Vd)}{\Hnorm_\KiA^2}\right] F^{(\KiA)} 
\eeq
where $F^{(\KiA)} = \int d\Vd \fB \HeHomo{}{\KiA}(\Vd)$.
Upon projecting along $\HeHomo{}{\KiB}(\Vd)$ we get
\beq\label{eq:Cherm}
\CHe{}{\KiB}=\int d\Vd \HeHomo{}{\KiB}(\Vd) [\CopC{\fB}] =
\sum_{\KiA=1}^{\Nherm}\left(\int d\Vd 
 \HeHomo{}{\KiB}(\Vd)\CopC{}\left[\weD(\Vd)
 \frac{\HeHomo{}{\KiA}(\Vd)}{\Hnorm_\KiA^2}\right]\right) F^{(\KiA)} =
\sum_{\KiA=1}^{\Nherm}C_{\KiB\KiA}F^{(\KiA)} 
\eeq
where the quantity in round brackets defines the elements $C_{\KiB\KiA}$ of
a $\Nherm\times\Nherm$ matrix $C$.

The third step is to express $F^{(\KiA)}$ in terms of the $\fBwe_\QiA$
\beq
\begin{split}
F^{(\KiA)} &= \int d\Vd \fB \HeHomo{}{\KiA}(\Vd) =
 \int d\Vd \weD(\Vd) \frac{\fB}{\weD(\Vd)} \HeHomo{}{\KiA}(\Vd) \\
&=\sum_\QiA \wi_\QiA \frac{\fB_i}{\weD(\Vd_\QiA)} \HeHomo{}{\KiA}(\Vd_\QiA)
=\sum_\QiA \fBwe_\QiA \HeHomo{}{\KiA}(\Vd_\QiA)
\end{split}
\eeq
where the last term can be rewritten in matrix notation as $H^T\fBwe$.

Combining the results obtained in the above three steps we can write
\eref{eq:ColVec} in matrix form
\beq
\fBwe'/\Dt=\wiMat H \Hnorm^{-2} C H^T \fBwe
\eeq
which identifies the collision matrix
\beq\label{eq:ColMat3}
\ColMat/\Dt=\wiMat H \Hnorm^{-2} C H^T = (H^T)^{-1,R} C H^T
\eeq 
For clarity we can equivalently write
$H^T_{\Nherm\Nvvec}\ColMat_{\Nvvec\Nvvec}=\Dt
C_{\Nherm\Nherm}H^T_{\Nherm\Nvvec}$ where the matrix dimensions are indicated by
explicit subscripts.

Eq.~\eref{eq:ColMat3} is a representation of the collision matrix that allows
its spectral analysis. Indeed the spectrum of $\ColMat$ is directly connected to
that of $C$.  In the square case $\Nherm=\Nvvec$ the two matrices are similar
and have the same spectrum. In the general rectangular case, since
$\Nvvec\geq\Nherm$, the spectrum of $C$ is contained in the spectrum of
$\ColMat$ (an eigenvector of $\ColMat$ being just $(H^T)^{-1,R}v$ where $v$ is
an eigenvector of $C$).  The additional $\Nvvec-\Nherm$ eigenvalues are just 0.

We have therefore reduced the problem to the computation of the spectrum of $C$.
We can deduce an explicit representation of this matrix using relations
\eref{eq:HcoeffC} and the defining equation \eref{eq:Cherm}. The matrix $C$ 
reads
\beq
\begin{pmatrix}
0 & & & \\
\begin{array}{|c|}
\hline
\\
(\aEFd) \\
\\
\hline
\end{array}
&
\begin{array}{|ccc|}
\hline
-\gFP & & \\
 & \ddots & \\
 & & -\gFP \\
\hline
\end{array} 
& & \\
& 
\begin{array}{|ccc|}
\hline
&& \\
&(\aEFd)& \\
&& \\
&& \\
\hline
\end{array}
&
\begin{array}{|cccc|}
\hline
-2\gFP & & & \\
 & \ddots & & \\
 & & \ddots & \\
 & & & -2\gFP \\
\hline
\end{array} 
& \\
& & & \ddots
\end{pmatrix}
\eeq
where $(\aEFd)$ contains only $\aEF_\DiA$ components, the square matrices are
diagonal, and all the remaining elements are zero.  The Fokker-Planck operator
fills the diagonal while $\CopB^{EXT}$ occupies the part below. The resulting
matrix is triangular and the eigenvalues are just given by the diagonal
elements, independently of the off-diagonal ones, i.e.
\beq
\lambda_\HiB = -\gFP \HiB \quad \HiB=0\ldots\Oherm
\eeq

Consequently, the collision matrix $\ColMat$ has eigenvalues $\lambda_\HiB\Dt$.
Going back to conditions~\eref{eq:stable}, the most stringent one is for
$\HiB=\Oherm$ and reads
\beq
0<\gFP\Dt<2/\Oherm
\eeq
which in the case of the $\Oherm=2$ scheme of Eq.~\eref{eq:LB2} reduces to
$0<\gFP\Dt<1$. These inequalities completely identify the range of model
parameters for which the scheme proposed in sec.~\ref{sec:hermite} does not lead
to an unbounded growth of the distribution functions with time.  Note that the
stability requirement imposes conditions only on the parameter $\gFP$
independently of the external field $\aEFd$. This is due to the initial
assumption that the field does not depend on the $\fBwe_\QiA$. Inclusion of
self-consistent force fields, that depend for example on the local density
Eq.~\eref{eq:LfmomE0}, would require a more careful analysis~\cite{WMS97}.

\section{Chapman-Enskog expansion}\label{sec:ChapEns}

A kinetic equation describes a system at the microscopic level of the
distribution function $\fBB{\Rd}{\Vd}{\ti}$.  Define the Knudsen number
$\epsilon$ as the ratio between the mean distance between two successive
particle collisions and the characteristic spatial scale of the system
(e.g. radius of an obstacle in a flow). If this number is very small the details
of particle collisions can be neglected and the system can be considered as a
continuum.  Using the Knudsen number as an expansion parameter, Chapman and
Enskog were able to derive from the Boltzmann equation the evolution of the
hydrodynamic variables (corresponding to the first moments of $\fB$) in the
continuum limit, thus reproducing the macroscopic Navier-Stokes
equations~\cite{ResiboisLeener}.  Eventually, the expansion has also been used
in the context of the LB method to derive the macroscopic equations obeyed by
Lattice Boltzmann models.  The fundamental hydrodynamic equations were recovered
consistently~\cite{RivetBoon}.

The Chapman-Enskog procedure is not restricted to the Boltzmann and
Lattice-Botzmann equations. In this section we apply it to the continuous
Fokker-Planck kinetic equation~\eref{eq:FPE2} and to the second-order
($\Oherm=2$) lattice scheme of Eq.~\eref{eq:LB2}.  We can then check if the same
macroscopic equations for the first moments are reproduced.

In the continuous case the expansion is straightforward. Indeed one can avoid it
completely and obtain the equations for the macroscopic variables by just
multiplying Eq.~\eref{eq:FPE2} by $ \Vb_{\DiA_1}\ldots\Vb_{\DiA_\KiB}$ and
integrating over velocity space.  In general at order $\KiB$, one obtains the
time-derivative of the $\KiB$-th moment plus the divergence of its flux on the
left-hand side.  On the right hand side the moments of $\Cop{\fB}$ can be
calculated using the Hermite expansion and the properties of Hermite
polynomials, as was done in the previous section to compute the collision
matrix.  Explicitly up to order one the result is
\begin{subequations}
\label{eq:macro1}
\bea
\partial_\ti \fBn + \partial_\DiA\fBJ_\DiA &=& 0 \label{eq:macro1a}\\
\partial_\ti \fBJ_\DiA + \partial_\DiB\fBP_{\DiA\DiB} 
 &=& -\gFP(\fBJ_\DiA-\fBn\uEF_\DiA) \label{eq:macro1b}
\eea
\end{subequations}
where we have introduced the \emph{external velocity}
$\uEF_\DiA=\aEF_\DiA/\gFP$.  The first is the continuity equation, the second
gives the evolution of the first moment $\fBJ_\DiA$, but involves also the
unknown second moment $\fBP_{\DiA\DiB}$. Indeed this procedure simply constructs
a non-closed hierarchy of equations for the moments of $\fB$. However, we are
not interested here in reproducing the Navier-Stokes equations, nor are we
interested in obtaining a closed set of equations. What we wish to check in the
following is whether the lattice scheme of sec.~\ref{sec:hermite} actually
reproduces the same hierarchy of equations.

In the discrete case we must make use of the complete Chapman-Enskog
expansion. To make it more transparent we have divided this derivation into
subsections.

\subsection{Preliminaries}

The macroscopic phenomena that we want to reproduce can occur on different time
and spatial scales. For example, there may be elastic effects, such as sound
propagation, with short time scales, and viscous effects, such as damping, with
longer time scales.  The idea of the Chapman-Enskog expansion is that assuming
such a separation of scales, these phenomena can be analyzed with multi-scale
asymptotic methods~\cite{BenderOrszag}.  We expand then the populations
$\fBwe_i$ and the spatial and time derivatives in powers of the parameter
$\epsilon$, the Knudsen number.  The hydrodynamic limit corresponds to $\epsilon
\ll 1$. In this limit, noticeable spatial variations take place typically over
distances of order $\epsilon^{-1}$. Hence, it is natural to introduce a
macroscopic space variable defined as $\Rd_1 =
\epsilon \Rd$.
If we expect to have both propagative and diffusive behavior, we must expand up
to second order in time, because in diffusion processes inhomogeneities at the
$\epsilon^{-1}$ space scale will relax on the $\epsilon^{-2}$ time
scale. Therefore we introduce two time variables $t_1=\epsilon t$ and
$t_2=\epsilon^{2} t$.  As usual in multi-scale methods, we then write
\begin{eqnarray}
\label{fexp}
\fBwe_i &=&\fBwe_i\so+\epsilon \fBwe_i\si+\epsilon^2 \fBwe_i\sii \\
\label{texp}
\dct&=&\epsilon\dt+\epsilon^2\ddt \\
\label{rexp}
\dca&=&\epsilon\da
\end{eqnarray}

Eq.~\eref{fexp} defines a corresponding expansion of the moments of
$\fBwe$ as
\beq\label{eq:exparho}
\fBn=\sum_\QiA \fBwe_\QiA 
= \sum_\QiA \left[\fBwe_i\so+\epsilon \fBwe_i\si+\epsilon^2 \fBwe_i\sii \right]
\doteq \fBn^{(0)}+\CEp\fBn^{(1)}+\CEp^2\fBn^{(2)} 
\eeq
and analogously for $\fBJ_\DiA, \fBP_{\DiA\DiB}$.  For convenience we also
rewrite the lattice collision operator Eq.~\eref{eq:LB2b} as
\bea\label{eq:LB2b.bis}
\LCopB[\fBwe_\QiA] &=& -\gFP\bar{\fBJ}_\DiA\frac{\Vb_{\QiA\DiA}}{\vT^2}\wi_\QiA
 -2\gFP\bar{\fBP}_{\DiA\DiB} 
\frac{\Vb_{\QiA\DiA}\Vb_{\QiA\DiB}-\vT^2\delta_{\DiA\DiB}}{2\vT^4}\wi_\QiA 
\eea
where
\bea
\bar{\fBJ}_\DiA &\doteq& \fBJ_\DiA-\fBn\uEF_\DiA \\
\bar{\fBP}_{\DiA\DiB} &\doteq& 
 \fBP_{\DiA\DiB}-\vT^2\fBn\delta_{\DiA\DiB}-
 \half(\uEF_\DiA\fBJ_\DiB+\uEF_\DiB\fBJ_\DiA)
\eea
Since $\bar{\fBJ}_\DiA$ and $\bar{\fBP}_{\DiA\DiB}$ depend linearly on the
moments $\fBn, \fBJ_\DiA, \fBP_{\DiA\DiB}$, we can write for them an expansion
similarly to Eq.~\eref{eq:exparho}.  Namely
$\bar{\fBJ}_\DiA=\bar{\fBJ}_\DiA^{(0)}+\epsilon\bar{\fBJ}_\DiA^{(1)}
+\epsilon^2\bar{\fBJ}_\DiA^{(2)}$ and analogously for $\bar{\fBP}_{\DiA\DiB}$.

\subsection{Expansion details}

The first step is to apply the expansions defined in the previous subsection to
both sides of Eq.~\eref{eq:LB2}.  On the left-hand side, we first manipulate the
streaming operator as usual in Chapman-Enskog expansions for lattice Boltzmann
models~\cite{WolfGladrow}.  Since the scale expansion parameter $\epsilon$ is
small, the populations vary little from one node the next.  We can approximate
the population $\fBweB{\QiA}{\Rd+\Vd_\QiA\Dt}{\ti+\Dt}$ by its Taylor expansion
around $\fBweB{\QiA}{\Rd}{\ti}$, and write up to second order in $\Dt$:
\begin{equation}\label{discretetime}
\fBweB{\QiA}{\Rd+\Vd_\QiA\Dt}{\ti+\Dt}-\fBweB{\QiA}{\Rd}{\ti} =
\Dt \left[ \dct + \Vb\ia\dca+
\frac{\Dt}{2}(\dct+\Vb\ia\dca)(\dct+\Vb\ib\dcb)\right]\fBweB{\QiA}{\Rd}{\ti}
\end{equation}
Using next Eqs.(\ref{discretetime}) and (\ref{fexp}-\ref{rexp}), the streaming
operator $[\fBweB{\QiA}{\Rd+\Vd_\QiA\Dt}{\ti+\Dt}-\fBweB{\QiA}{\Rd}{\ti}]/\Dt$
can be expanded in powers of $\epsilon$ as:
\begin{eqnarray}
\label{stream0}
{\rm order} \; \epsilon^0 &:& 0 \\
\label{stream1}
{\rm order} \; \epsilon^1 &:& [\dt+\Vb\ia\da]\fBwe_i\so\\
\label{stream2}
{\rm order} \; \epsilon^2 &:& [\dt+\Vb\ia\da]\fBwe_i\si
+[\ddt+\frac{\Delta t}{2}(\dt+\Vb\ia\da)(\dt+\Vb\ib\db)]\fBwe_i\so
\end{eqnarray}
On the right hand side, in the case of the lattice collision operator 
the expansion acts order by order on the moments and we can write
\bea
\LCopB &=& \LCopB\so+\LCopB\si+\LCopB\sii
\eea
where
\bea
\LCopB^{(\HiB)}
&=& -\gFP\bar{\fBJ}^{(\HiB)}_\DiA\frac{\Vb_{\QiA\DiA}}{\vT^2}
 \wi_\QiA -2\gFP\bar{\fBP}^{(n)}_{\DiA\DiB} 
\frac{\Vb_{\QiA\DiA}\Vb_{\QiA\DiB}-\vT^2\delta_{\DiA\DiB}}{2\vT^4}\wi_\QiA 
\eea
for $\HiB=0,1$ and $2$.

The second step is to equate corresponding orders of the expansion. Thus
we obtain to order $\CEp^0$
\beq\label{eq:CE0}
0=-\gFP\bar{\fBJ}\so_\DiA\frac{\Vb_{\QiA\DiA}}{\vT^2}\wi_\QiA 
 -2\gFP\bar{\fBP}\so_{\DiA\DiB} 
\frac{\Vb_{\QiA\DiA}\Vb_{\QiA\DiB}-\vT^2\delta_{\DiA\DiB}}{2\vT^4}\wi_\QiA
\eeq
to order $\CEp^1$ 
\beq\label{eq:CE1}
\partial_\ti^{(1)}\fBwe_{\QiA}^{(0)}
 +\Vb_{\QiA\DiA}\partial_\DiA^{(1)}\fBwe_{\QiA}^{(0)}
=-\gFP\bar{\fBJ}\si_\DiA\frac{\Vb_{\QiA\DiA}}{\vT^2}\wi_\QiA
 -2\gFP\bar{\fBP}_{\DiA\DiB}^{(1)}
 \frac{\Vb_{\QiA\DiA}\Vb_{\QiA\DiB}-\vT^2\delta_{\DiA\DiB}}{2\vT^4}\wi_\QiA
\eeq
and to order $\CEp^2$ the equation can be rewritten more conveniently as
\bea\label{eq:CE2}
&&\partial_\ti^{(1)}\fBwe_{\QiA}^{(1)}+
\partial_\DiA^{(1)}\Vb_\DiA\fBwe_{\QiA}^{(1)}+
\partial_\ti^{(2)}\fBwe_{\QiA}^{(0)}+
\frac{\Dt}{2}[
\partial_\ti^{(1)}(
 \partial_\ti^{(1)}\fBwe_{\QiA}^{(0)}+
 \partial_\DiB^{(1)}\Vb_\DiB\fBwe_{\QiA}^{(0)})+
 \nonumber \\
&& \quad
\partial_\DiA^{(1)}(
 \partial_\ti^{(1)}\Vb_\DiA\fBwe_{\QiA}^{(0)}+
 \partial_\DiB^{(1)}\Vb_\DiA\Vb_\DiB\fBwe_{\QiA}^{(0)})] = 
 -\gFP\bar{\fBJ}\sii_\DiA\frac{\Vb_{\QiA\DiA}}{\vT^2}\wi_\QiA
-2\gFP\bar{\fBP}_{\DiA\DiB}^{(2)}
 \frac{\Vb_{\QiA\DiA}\Vb_{\QiA\DiB}-\vT^2\delta_{\DiA\DiB}}{2\vT^4}\wi_\QiA
\eea

The third step is to compute the moment equations associated with
Eqs.~\eref{eq:CE0}-\eref{eq:CE1}.  For the zeroth moment equation one can just
sum both sides of the equations over $i$, for the next moments one must first
multiply by $\Vb_\DiC$, $\Vb_\DiC\Vb_\DiD$, and so on. Note that the orders of
the velocity moments are not the orders of the $\epsilon$-expansion. For each
order in $\epsilon$ we can compute different moment equations.  As we will show
shortly, for the purpose of reproducing the macroscopic
equations~\eref{eq:macro1} we need up to the second moment equation for orders
$\epsilon^0$ and $\epsilon^1$ and only to the first moment equation for order
$\epsilon^2$. The computations are carried using the relations of
appendix~\ref{sec:hermitepoly}.  To order $\epsilon^0$, the zeroth moment does
not give any information, the first and second moment read
\begin{subequations}
\bea
0 &=& -\gFP\bar{\fBJ_\DiC}\so \label{eq:lFP.0b} \\
0 &=& -2\gFP \bar{\fBP}_{\DiC\DiD}^{(0)} \label{eq:lFP.0c}
\eea
\end{subequations}
To order $\epsilon^1$ we find for the zeroth, first and second moments:
\begin{subequations}
\bea
\partial_\ti^{(1)}\fBn^{(0)}+\partial_\DiA^{(1)}\fBJ_\DiA^{(0)} 
 &=& 0 \label{eq:lFP.1a}\\
\partial_\ti^{(1)}\fBJ_\DiC^{(0)}+\partial_\DiA^{(1)}\fBP_{\DiA\DiC}^{(0)} 
 &=& -\gFP\bar{\fBJ}\si_\DiC \label{eq:lFP.1b} \\
\partial_\ti^{(1)}\fBP_{\DiC\DiD}^{(0)}
 +\partial_\DiA^{(1)}\fBQ_{\DiA\DiC\DiD}^{(0)} 
 &=&  -2\gFP \bar{\fBP}_{\DiC\DiD}^{(1)} \label{eq:lFP.1c}
\eea
\end{subequations}
And to order $\epsilon^2$ we only consider the zeroth and first moment equations
\begin{subequations}
\bea
\partial_\ti^{(2)}\fBn^{(0)}+\partial_\ti^{(1)}\fBn^{(1)}
 +\partial_\DiA^{(1)}\fBJ_\DiA^{(1)}
 -\frac{\gFP\Dt}{2}\partial_{\DiA}^{(1)}\bar{\fBJ}_\DiA^{(1)} &=& 0 
 \label{eq:lFP.2a} \\
\partial_\ti^{(2)}\fBJ_\DiC^{(0)}+\partial_\ti^{(1)}\fBJ_\DiC^{(1)}+
 \partial_\DiA^{(1)}\fBP_{\DiA\DiC}^{(1)}+\frac{\Dt}{2}[
 \partial_\ti^{(1)}(-\gFP\bar{\fBJ}\si_\DiC)+
 \partial_{\DiA}^{(1)}(-2\gFP \bar{\fBP}_{\DiA\DiC}^{(1)})] 
&=& -\gFP\bar{\fBJ}\sii_\DiC
 \label{eq:lFP.2b}
\eea
\end{subequations}
where we made use of \eref{eq:lFP.1a}, \eref{eq:lFP.1b} to derive the first, 
and of \eref{eq:lFP.1b}, \eref{eq:lFP.1c} for the second equation.

\subsection{Macroscopic equations}

We can now add up the equations at different orders in $\epsilon$ and obtain
expanded macroscopic equations for the zeroth and first moments of the
populations $\fBwe_\QiA$.  The final step then is to reconstruct the derivative
operators from the expanded ones.

For the zeroth moment, we construct 
$\CEp^1\cdot$\eref{eq:lFP.1a}$+\CEp^2\cdot$\eref{eq:lFP.2a}
(order $\CEp^0$ does not add anything in this case)
and we get
\beq\label{eq:lhydro0}
\CEp\partial_\ti^{(1)}\fBn^{(0)}+\CEp\partial_\DiA^{(1)}\fBJ_\DiA^{(0)}
+\CEp^2\partial_\ti^{(2)}\fBn^{(0)}+\CEp^2\partial_\ti^{(1)}\fBn^{(1)}+
\CEp^2\partial_\DiA^{(1)}\fBJ_\DiA^{(1)}
-\CEp^2\frac{\gFP\Dt}{2}\partial_{\DiA}^{(1)}\bar{\fBJ}_\DiA^{(1)}=0 
\eeq
For the first moment, we construct
$\CEp^0\cdot$\eref{eq:lFP.0b}$+\CEp^1\cdot$\eref{eq:lFP.1b}
$+\CEp^2\cdot$\eref{eq:lFP.2b}
and we obtain
\bea\label{eq:lhydro1}
&& \CEp\partial_\ti^{(1)}\fBJ_\DiC^{(0)}
 +\CEp\partial_\DiA^{(1)}\fBP_{\DiA\DiC}^{(0)}
 +\CEp^2\partial_\ti^{(2)}\fBJ_\DiC^{(0)}
 +\CEp^2\partial_\ti^{(1)}\fBJ_\DiC^{(1)}
 +\CEp^2\partial_\DiA^{(1)}\fBP_{\DiA\DiC}^{(1)}+ \nonumber \\
&&\quad
 \CEp^2\frac{\Dt}{2}[
 \partial_\ti^{(1)}(-\gFP\bar{\fBJ}\si_\DiC)+
 \partial_{\DiA}^{(1)}(-2\gFP \bar{\fBP}_{\DiA\DiC}^{(1)})]=
 -\CEp^1\gFP\bar{\fBJ}\si_\DiC-\CEp^2\gFP\bar{\fBJ}\sii_\DiC
\eea

In these equations both the moments of $\fBwe_\QiA$ (corresponding to the
macroscopic variables) and the differential operators are expanded up to order
$\epsilon^2$. We can straightforwardly reconstruct the original quantities using
relation \eref{eq:exparho} and the analogous ones for the other variables.  The
spatial derivative is reconstructed in the same spirit noting that
$\partial_\DiA X = \CEp\partial_\DiA^{(1)} (X^{(0)}+\CEp X^{(1)}) =
\CEp\partial_\DiA^{(1)} X^{(0)}+\CEp^2 \partial_\DiA^{(1)} X^{(1)}$, 
where $X$ is
any of the moments.  In a similar fashion, for the time derivatives
$\partial_\ti X =
\CEp\partial_\ti^{(1)}X^{(0)}+\CEp^2\partial_\ti^{(2)}X^{(0)}+\CEp^2
\partial_\ti^{(1)}X^{(1)}$, where a term of order $\CEp^3$ was omitted.
Inserting Eqs.\eref{eq:lFP.0b},\eref{eq:lFP.0c} where necessary, equations
\eref{eq:lhydro0}, \eref{eq:lhydro1} can then be rewritten as
\begin{subequations}\label{eq:CE}
\begin{eqnarray}
\dct \rho + \dca J_\alpha
&=& \frac{\gamma\Delta t}{2}
\dca (\fBJ_\DiA-\fBn\uEF_\DiA) \\
\nonumber
\dct J_\alpha + \dcb \fBP_{\alpha\beta}  
&=&  -\gFP(\fBJ_\DiA-\fBn\uEF_\DiA) \\
\nonumber
&\;& \;\;  + \, \gamma\Delta t \, \dcb \left(
\fBP_{\DiA\DiB}-\vT^2\fBn\delta_{\DiA\DiB}
 -\half(\uEF_\DiA\fBJ_\DiB+\uEF_\DiB\fBJ_\DiA)
 \right) \\
&\;& \;\;  + \, \frac{\gamma\Delta t}{2}  \, \dct \left( 
\fBJ_\DiA-\fBn\uEF_\DiA\right) 
\end{eqnarray}
\end{subequations}
Interestingly, we find that these equations differ from the continuous equations
by one additional term in the first and by two terms in the second. All
corrections are of order $\gamma\Delta t$.

We can gain more insight in these results by rewriting them in a slightly
different way.  Let $\fBwe_\QiA^{eq}$ be the solutions of
$\LCopB[\fBwe_\QiA]=0$. From the explicit form \eref{eq:LB2b.bis} we see that
the $\fBwe_\QiA^{eq}$ satisfy $\bar{\fBJ}_\DiA=\bar{\fBP}_{\DiA\DiB}=0$, or
equivalently
\bea
\fBJ_\DiA&=&\fBn\uEF_\DiA\doteq\fBJ_\DiA^{eq} \label{eq:eqJ} \\
\fBP_{\DiA\DiB}&=&\vT^2\fBn\delta_{\DiA\DiB}
 +\half(\uEF_\DiA\fBJ_\DiB+\uEF_\DiB\fBJ_\DiA)
\doteq\fBP_{\DiA\DiB}^{eq} \label{eq:eqP}
\eea
With these definitions we can rewrite Eqs.\eref{eq:CE} as
\begin{subequations}\label{eq:CEeq}
\begin{eqnarray}
\label{cfprhoeveq}
\dct \rho + \dca J_\alpha
&=& \frac{\gamma\Delta t}{2}
\dca (J_\alpha - J_\alpha^{eq}) \\
\nonumber
\dct J_\alpha + \dcb \fBP_{\alpha\beta}  
&=&  -\gamma (J_\alpha -J_\alpha^{eq}) \\
\nonumber
&\;& \;\;  + \, \gamma\Delta t \, \dcb \left(\fBP_{\alpha\beta}-
\fBP_{\alpha\beta}^{eq} \right) \\
\label{cfpjev2eq}
&\;& \;\;  + \, \frac{\gamma\Delta t}{2}  \, \dct \left( 
J_\alpha - J_\alpha^{eq}\right) 
\end{eqnarray}
\end{subequations}
This form shows that for a given value of $\gamma$, the closer to equilibrium
the system is, the closer the evolution of the discrete system is to that of the
continuous Fokker-Planck equation.

Eqs.~\eref{eq:CEeq} are the final outcome of the Chapman-Enskog expansion of the
numerical scheme of Eqs.~\eref{eq:LB2}. Together with the stability results of
sec.\ref{sec:colli}, they complete the analysis of the proposed numerical
method. Unfortunately, they prove that the scheme does not actually solve the
continuous kinetic equation~\eref{eq:FPE2}, because of the additional terms in
Eqs.~\eref{eq:CEeq} with respect to Eqs.~\eref{eq:macro1}. However, as
just illustrated, we know explicitly the error made. In the next section
we exploit this knowledge to build a corrected scheme that is able
to solve the continuous equation.

\section{Lattice Fokker-Planck Algorithm}\label{sec:corrected}\label{sec:algo}

Having in mind the results of the previous section we provide here a correct
numerical procedure to solve the continuous Fokker-Planck kinetic
equation~\eref{eq:FPE1}. 

The results of the Chapman-Enskog expansion suggest that by properly redefining
the hydrodynamic variables it is possible to recover the correct continuous
macroscopic equations. Let
\begin{subequations}\label{defjPst}
\bea
J_\alpha\st &=& \left(1-\frac{\gamma\Delta t}{2}\right) J_\alpha 
+\frac{\gamma\Delta t}{2} J_\alpha^{eq} \\
\fBP_{\DiA\DiB}\st &=& \left(1-\gamma\Delta t\right) \fBP_{\alpha\beta}
+ \gamma\Delta t \fBP_{\alpha\beta}^{eq} 
\eea
\end{subequations}
Then Eqs.\eref{eq:CEeq} are rewritten as
\begin{subequations}\label{eq:FPEsim}
\bea
\dct \fBn + \dca \fBJ_\DiA\st &=& 0 \\
\dct \fBJ_\DiA\st + \dcb \fBP_{\alpha\beta}\st  
&=&  -\tilde{\gamma} (\fBJ_\DiA\st -\fBn \uEF_\DiA) 
\eea
\end{subequations}
where an effective friction 
\begin{equation}\label{eq:gammatilde}
\frac{1}{\tilde{\gamma}} = \frac{1}{\gamma} - \frac{\Delta t}{2}
\end{equation}
is introduced.  At the level of the Chapman-Enskog expansion the above equations
correspond to the continuous Eqs.~\eref{eq:macro1}.  A similar approach was used
in~\cite{GZS02,LV01} by redefinition of velocity in the presence of a forcing
term.  The quantities $J_\alpha\st$ and $\fBP_{\DiA\DiB}\st$ reduce to
$\fBJ_\DiA$ and $\fBP_{\DiA\DiB}$ in the limit $\gFP\Dt \rightarrow 0$,
Furthermore, if $J_\alpha= J_\alpha^{eq}=\rho \uEF_\DiA$ we also have
$J_\alpha\st=J_\alpha= J_\alpha^{eq}$ (and the same for the stress tensor). With
walls these equalities do not hold in general. Indeed, the boundary conditions
set $J_\alpha\st=0$, whereas $J_\alpha^{eq}$ is non-zero when a field is
applied.

A computational algorithm to solve \eref{eq:FPE1} can be divided
into two parts, the first initializes the simulation, and the
second is the dynamical evolution of the $\fBwe_\QiA$.
\begin{description}
\item[Initialization.] If we perform a simulation using the \qmark{bare} 
definition of the moments in the collision operator, but sample the
\qmark{corrected} moments $\rho$, $J_\alpha\st$ and $\fBP_{\alpha\beta}\st$, the
latter satisfy the continuous Fokker-Planck equation with second-order accuracy,
see Eq.~\eref{eq:FPEsim}, but with a rescaled friction $\tilde{\gamma}$.
Suppose we want to simulate a system with a friction $\gFP_0$. Then in
Eq.~\eref{eq:gammatilde} we identify $\tilde{\gFP}$ with $\gFP_0$, solve for
$\gFP$ obtaining
\begin{equation}\label{eq:gamma0}
\gamma = \frac{\gamma_0}{1+\displaystyle\frac{\gamma_0 \Delta t}{2} }
\end{equation} 
and use this $\gFP$ in the simulation. The external velocity $\uEF_\DiA$ must
remain unaffected. So if one wants to apply a field $\aEF_{0,\DiA}$, one must
use in the simulation a field $\aEF_\DiA$ such that
$\aEF_\DiA/\gFP=\aEF_{0,\DiA}/\gFP_0$.  The initial conditions are set on the
starred variables, defined by \eref{defjPst} using $\gamma$, not $\gamma_0$.
\item[Simulation loop.]
Given the set of $\fBweB{\QiA}{\Rd}{\ti}$ at time $\ti$, the
$\fBweB{\QiA}{\Rd}{\ti+\Dt}$ at time $\ti+\Dt$ are found, for each $\Rd$, by the
following steps
\ben
\item compute the moments of $\fB$ using \eref{eq:Lfmom}, or explicitly
 \eref{eq:LfmomE}
\item compute the Hermite coefficients of $\fB$, Eq.\eref{eq:HcoeffF}
\item compute the Hermite coefficients of $\Cop{\fB}$, Eq.\eref{eq:HcoeffC}
\item compute the right-hand side of \eref{eq:LB}
\item compute the left-hand side of \eref{eq:LB}
\een
Then the procedure is repeated. At regular times we can sample the hydrodynamic
observables of interest corresponding to the moments of $\fB$. One has to take
care however to sample the starred variables, because those are the ones that
correctly reproduce the continuous equations.
\end{description}
An extensive literature is available for the implementation of the Lattice
Boltzmann method, where important issues such as boundary conditions and large
scale code optimization have been investigated in depth~\cite{Succi}.  Most of
the LB techniques can be directly extended to the Lattice Fokker-Planck method.
For further details we advise then the interested reader to more specialized
articles, such as the performance studies of~\cite{Donath03,dimeKlausBach03}.

\section{Numerical results}\label{sec:numerics}

\subsection{Numerical limits}

Combining the results of secs.~\ref{sec:colli} and \ref{sec:ChapEns}, one finds
that the second-order scheme of the previous section allows one to solve the FP
equation for $0<\gFP\Dt<1$, independently of the external field $\aEFd$, and the
smaller $\gFP\Dt$, the closer the lattice solution will be to the continuous
one.  Moreover, since in the numerical scheme we use the friction $\gFP$ given
by Eq.~\eref{eq:gamma0} instead of the real $\gFP_0$, the stability condition
actually corresponds to $0<\gamma_0\Dt<2$, so that it seems possible to simulate
systems with a time step longer than the reciprocal friction. These theoretical
findings need some numerical back-up.  To this purpose we consider here two
basic examples in $\Dim=1$ for which analytical solutions are also available and
we compare these solutions with the outcome of simulations in the D1Q3
lattice~\cite{Succi}.  We can then check the validity of the proposed scheme and
set constraints on the range of parameters.

\begin{figure}
\begin{center}
\includegraphics[]{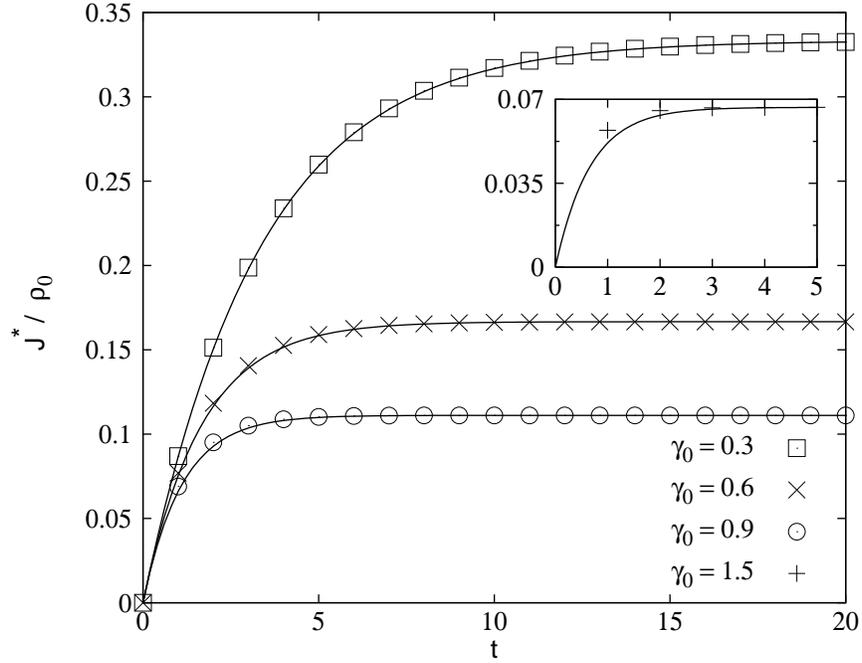} 
\caption{
A constant field is applied by imposing an external acceleration
$\aEF_0=0.1$. As a consequence a flux $J^*(\ti)$ is created in an initially
homogeneous system at density $\rho_0$.  Because of friction forces, the
quantity $J^*/\rho_0$ saturates at the steady state value $\aEF_0/\gFP_0$.  Time
$\ti$ is in units of $\Dt$ and the vertical axis has units of $\Delta x/\Dt$
where $\Delta x$ is the lattice spacing. The algorithm reproduces the continuous
solution (solid lines), but only in the range $0<\gFP_0\Dt<1$.  For
$\gFP_0\Dt>1$ (inset) discrepancy with the analytical solution is found.}
\label{figgammaeffcook}
\end{center}
\end{figure}

In the first example, we consider a system with periodic boundary conditions,
initially homogeneous at density $\rho_0$, and with no initial velocity.  A
constant homogeneous field is applied resulting in an external acceleration
$\aEF_0$.  From the solution of the continuous equations~\eref{eq:macro1}, the
density does not evolve, while the flux $J$ is uniform and evolves as
\begin{equation}
J = (\rho_0 \aEF_0/\gFP_0) \left(1 - e^{-\gamma_0 t}\right)
\end{equation}
Direct simulation of the system without the $\tilde{\gFP}$ prescription of
sec.~\ref{sec:algo} leads to an exponential solution with a wrong
rate. Including the prescription and sampling the starred moment $\fBJ\st$, one
obtains the correct result, as shown in \fig{figgammaeffcook}.  However, for
$\gamma_0 \Dt \ge 1$ ($\gamma \Dt \ge 2/3$), we find that the numerical results
deviate from the continuous solution, so that, even if possible in principle, it
is necessary in practice to constrain also the friction $\gFP_0$ to the range
$0<\gFP_0\Dt<1$.

\begin{figure}
\begin{center}
\includegraphics[]{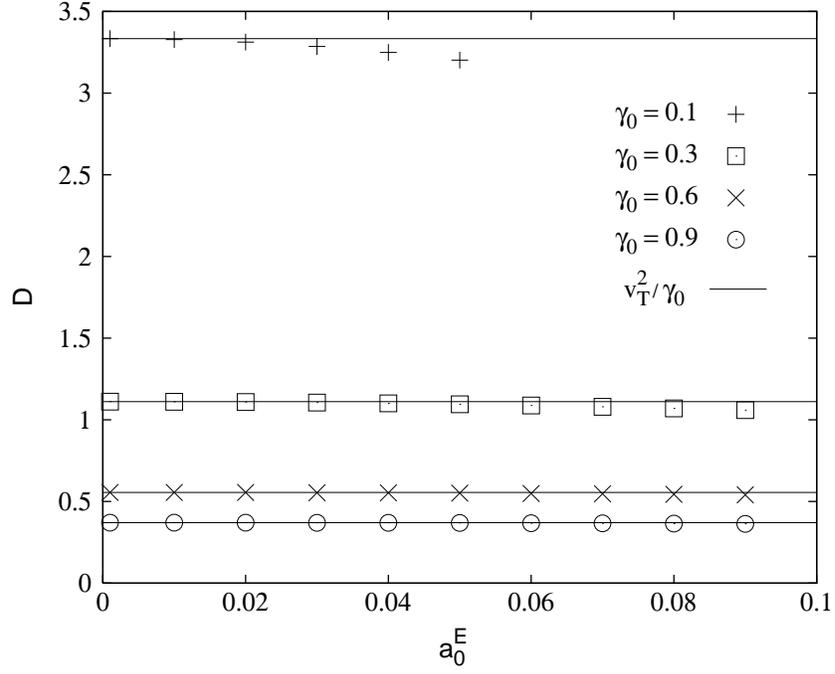}
\caption{
A constant external acceleration $\aEF_0$ is applied in a confined system 
initially homogeneous. The density converges to the barometric law from
which a simulated diffusion coefficient $\Dsim$ can be extracted.
The continuous lines correspond to Einstein's relation. The deviation
are caused by the discretized solution and can be explained with
the results of the Chapman-Enskog expansion. The acceleration is in
units $\Delta x/\Delta t^2$ and the diffusion coefficients in units
$\Delta x^2/\Delta t$, where $\Delta x$ is the lattice spacing.}
\label{figdeff}
\end{center}
\end{figure}

In the second example, we consider the same system, but with bounce-back no-slip
reflecting boundary conditions~\cite{Succi}.  A constant field is applied
resulting in an external acceleration $\aEF_0$.  Accumulation due to migration
results in a concentration gradient which is the source of a diffusive flux
opposed to the applied field.  From the balance of fluxes, we find at
equilibrium the barometric law for the density
\beq\label{eq:barolaw}
\rho(x)\propto\exp\left(\frac{\aEF_0}{\vT^2} x\right)
\eeq
where $\vT^2=k_B T/m$ is the thermal velocity.  The same result is obtained from
the direct solution of Eqs.\eref{eq:macro1} using the assumption that the tensor
$\fBP_{\DiA\DiB}$ has already relaxed to its equilibrium value
$\fBP_{\DiA\DiB}^{eq}$ given by Eq.~\eref{eq:eqP}.  Simulations without the
$\tilde{\gFP}$ prescription give an exponential profile, but the exponential
slope wrongly shows a dependence on the friction $\gFP_0$.  With the correct
prescription the profile is still exponential, and in order to check the slope,
we first rewrite the fraction in the right hand side of~\eref{eq:barolaw} as
$\aEF_0/(\gFP_0 \Dtheo)$ where, from Einstein's relation, $\Dtheo=\vT^2/\gFP_0$.
From an exponential fit of our data we can then derive a simulated diffusion
coefficient $\Dsim$ dividing $\aEF_0/\gFP_0$ by the slope.  The numerical
results are reported in \fig{figdeff} compared to the continuous value $\Dtheo$.
Slight deviations can be observed, especially for small $\gFP_0$.
These findings can be explained using the outcome of the Chapman-Enskog
analysis. At steady state equations~\eref{eq:FPEsim} become
\bea
\partial_\DiA \fBJ_\DiA\st &=& 0 \\
\partial_\DiB \fBP_{\DiA\DiB}\st&=& -\tilde{\gFP}(\fBJ_\DiA\st-\fBn\uEF_\DiA)
\eea
Using the assumption $\fBP_{\DiA\DiB}\st=\fBP_{\DiA\DiB}^{eq}=
\vT^2\fBn\delta_{\DiA\DiB}+\half(\uEF_\DiA\fBJ_\DiB+\uEF_\DiB\fBJ_\DiA)$, and
and the definition of $\fBJ_\DiA\st$, Eq.~\eref{defjPst}, we arrive at the
equations
\bea
\partial_\DiA \fBJ_\DiA\st &=& 0 \\
\vT^2\partial_\DiA\fBn+\frac{1}{2-\gFP\Dt} 
 [-\gFP\Dt\uEF_\DiA\uEF_\DiB\partial_\DiB\fBn
  +\uEF_\DiB\partial_\DiB\fBJ_\DiA\st]
&=& -\tilde{\gFP} (\fBJ_\DiA\st-\fBn\uEF_\DiA)\label{eq:diff3}
\eea
for the redefined variables $\fBn,\fBJ_\DiA\st$. Note that in this case
$\tilde{\gFP}$ must be identified with $\gFP_0$ above. Using the equilibrium
result $\fBJ_\DiA\st=0$, Eq.~\eref{eq:diff3} easily yields an exponential
solution for $\fBn(x)$ in dimension $\Dim=1$, from which we can derive the
simulated diffusion coefficient $\Dsim$ as
\begin{equation}\label{eq:Dcorr}
\Dsim=\frac{\vT^2}{\gamma_0}-\frac{\Dt}{2\gFP_0^2}(\aEF_0)^2
\end{equation}
The first term is Einstein's relation and the second gives a correction which is
small for vanishing external fields.  The result in Eq.~\eref{eq:Dcorr} is in
accordance with the values reported in \fig{figdeff} since the correction is
larger for small $\gFP_0$. Eq.~\eref{eq:Dcorr} can also be written as
\begin{equation}
\Dsim=\Dtheo
 \left[1-\frac{\gFP_0\Dt}{2}\left(\frac{\uEF}{\vT}\right)^2\right]
\end{equation}
where $\uEF=\aEF_0/\gFP_0$. Then another way of interpreting the correction is
to say that our numerical scheme is more valid in a low Mach number
regime, i.e. $\uEF$ must be small compared to $\vT$, which numerically
is $1/\sqrt{3}\simeq 0.6$ for D1Q3 and most common lattices.

Summarizing, we have found that the scheme works, but the theoretical range of
parameters must be restrained.  The physical friction $\gFP_0$ which can be
simulated must be such that $0<\gFP_0\Dt<1$.  A small $\gFP_0$ is good to obtain
a discrete evolution closer to the continuous one, but it must not be too small
compared to the external acceleration $\aEF_0$ since otherwise the low Mach
number assumption would fail. As a final remark, in the case of spatially
dependent forces, this must be true for all the points of the system, as we show
in the next section.

\subsection{Further examples}

In this section we apply the LFP method to two 
cases for which no obvious analytical solutions are
available.

\begin{figure}
\begin{center}
\includegraphics[width=\textwidth]{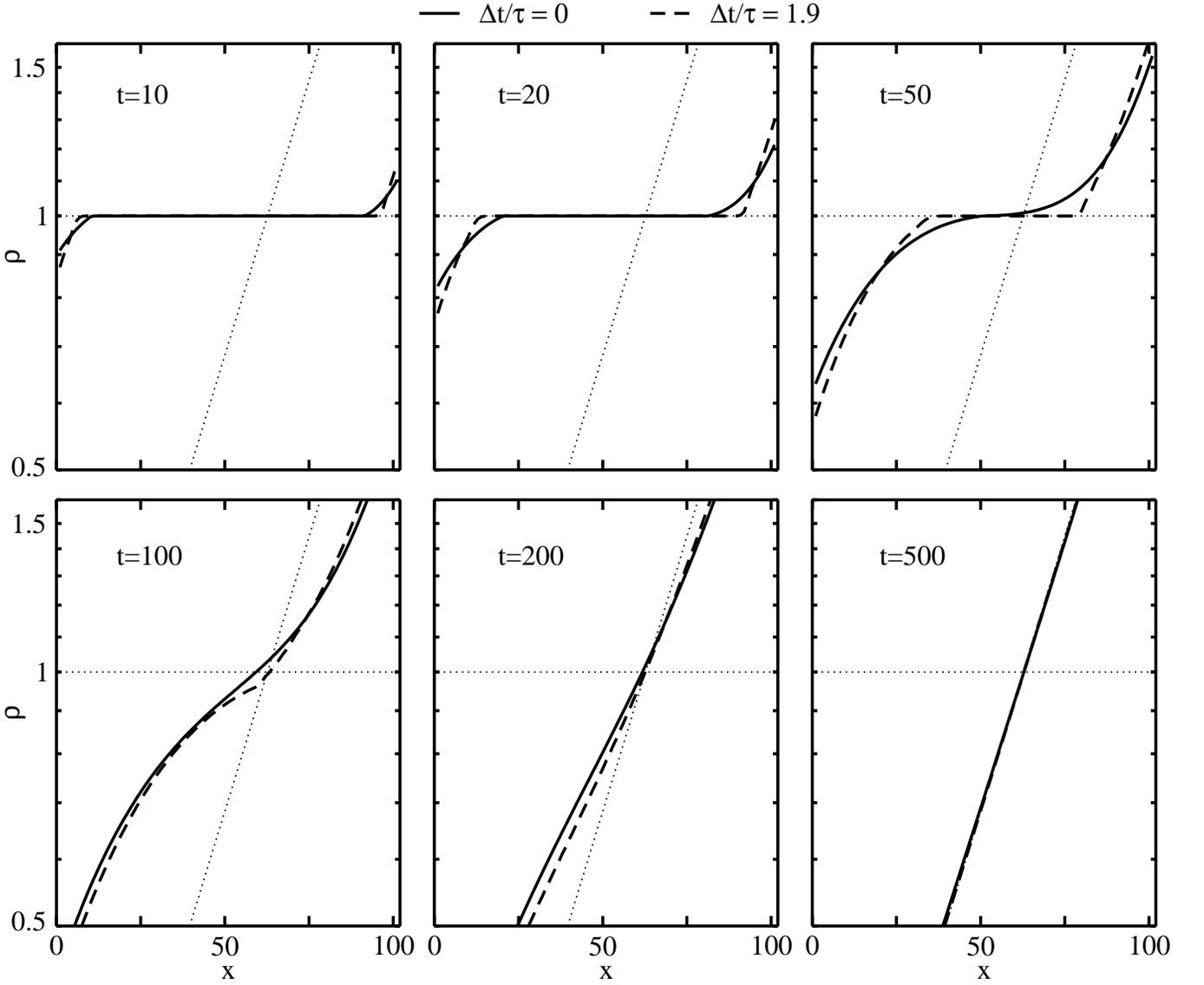}
\caption{Log-linear graphs of the density $\fBn(x)$ as function of the
position $x$ at six different times for a 1$\Dim$ confined system with friction
$\gFP_0=0.05$ under the influence of an external acceleration $\aEF_0=0.01$.  On
a D1Q3 grid of 101 points, the system is initially prepared at a homogeneous
density $\fBn_0=1$ (horizontal dotted line in the figures) and no initial
velocity.  As the field is applied the system gradually evolves to the
barometric law, diagonal dotted line in the figures. We show the evolution for a
pure FP collision operator ($\tauinvDt=0$, full line) and combined with a BGK
operator ($\tauinvDt=1.9$, dashed line).}
\label{fig:trans1d}
\end{center}
\end{figure}

In the first case, consider the 1$\Dim$ system of the previous subsection with
reflecting boundaries and simulations on the D1Q3 lattice.  We focus here on the
time-dependent approach to equilibrium.  Such a condition corresponds to
sedimentation caused by gravity. We also combine the FP collision operator with
a BGK operator with a single relaxation time $\tau$ to account for collisions
between particles, and possibly hydrodynamic interactions.  The stability
analysis of sec.~\ref{sec:colli} can be carried straightforwardly also in this
combined case, giving the constraint $0<2\gFP\Dt+\tauinvDt<2$, where $\gFP\Dt$
in our scheme is given by Eq.~\eref{eq:gamma0}. Given a value of $\gFP_0$ in
accordance to the previous subsection, we can then afford a value of $\tauinvDt$
up to $2-2\gFP_0\Dt$ and slightly above.  We report the results in
\fig{fig:trans1d}. Interestingly we find that the presence of BGK collisions
delays the start of the relaxation, but then makes it converge more quickly once
started.

\begin{figure}
\begin{center}
\includegraphics[]{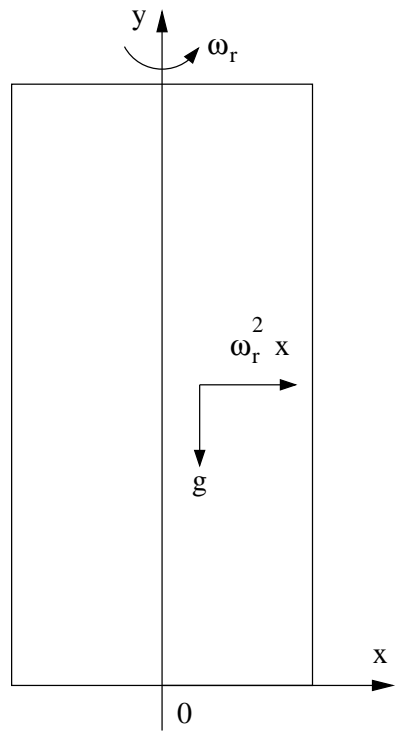}
\caption{The 2$\Dim$ system. Sedimentation under gravity
is combined with a centrifugal force. The external acceleration
is a combination of the constant $g$ and the linear escape term
$\omega_r^2 x$.}
\label{fig:centrifugal}
\end{center}
\end{figure}

\begin{figure}
\begin{center}
\includegraphics[width=\textwidth]{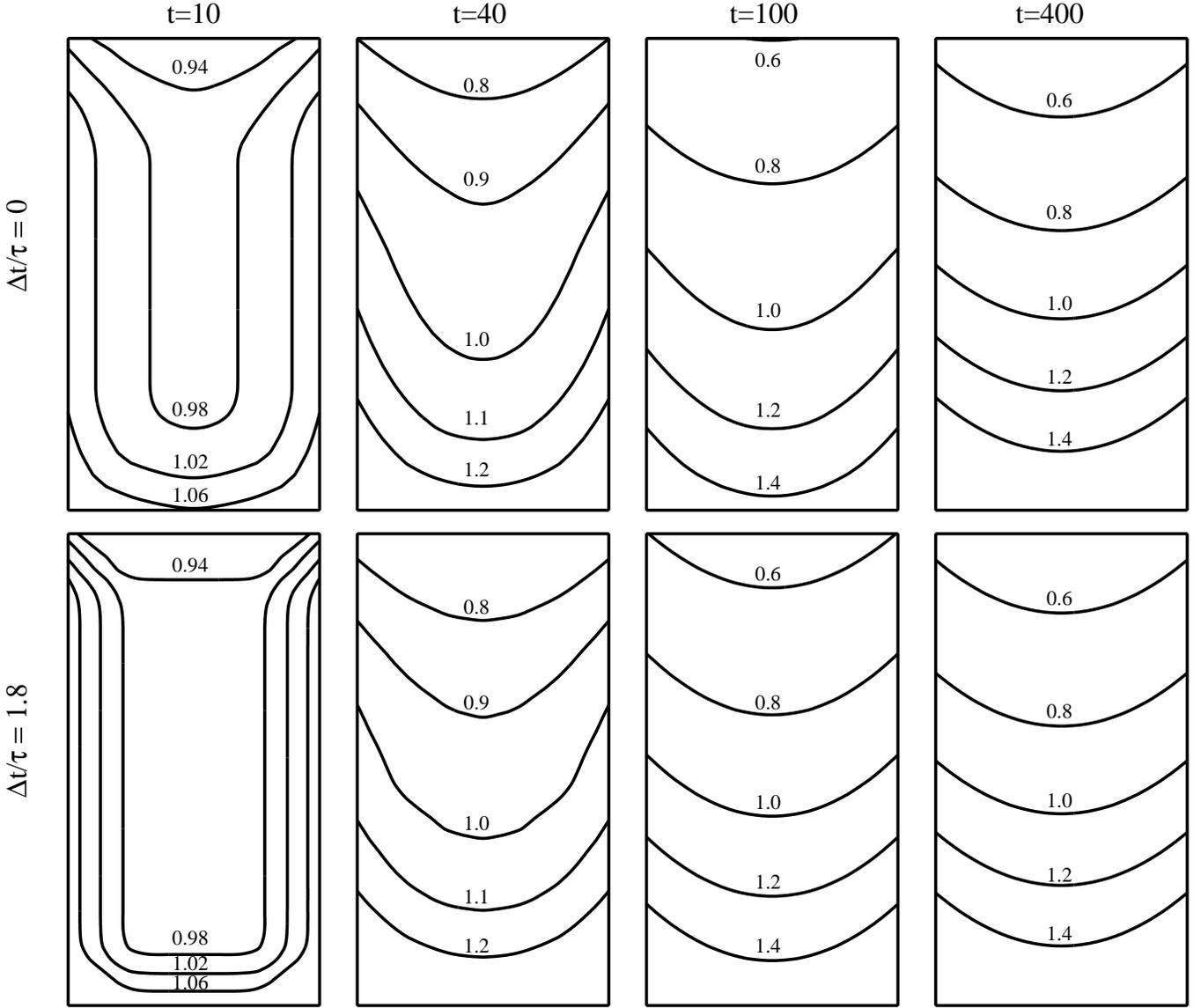}
\caption{Contour plots of the density $\fBn(x,y)$ 
at four different times.  The system is initially homogeneous at density
$\fBn_0=1$ and no initial velocity.  As the field is applied the density
gradually evolves to the characteristic parabolic profiles of the equilibrium
distribution. We show the evolution for a pure FP collision operator
($\tauinvDt=0$) and combined with a BGK operator ($\tauinvDt=1.8$). 
At $\ti=400$ both systems are converged to the analytical Boltzmann law.}
\label{fig:trans2d}
\end{center}
\end{figure}

In the second case, we consider the 2$\Dim$ system represented in 
\fig{fig:centrifugal}, where we combine sedimentation and a centrifugal
force. 
We apply bounce-back reflecting boundaries on a D2Q9 lattice~\cite{Succi} of
21$\times$41 points. We consider a system with friction $\gFP_0=0.1$ under the
influence of a gravity $g=0.01$ and a centrifugal force due to a rotation of
frequency $\omega_r=0.03$.  Also at the $x$ borders the low-Mach assumption is
satisfied.  We report the results in \fig{fig:trans2d}.  At short times the pure
FP system departs earlier from the homogeneous situation. However, at longer
times the presence of BGK collisions makes the system approach faster the
equilibrium distribution. Around $\ti=400$ both systems are converged and the
final profiles are in agreement with the analytical Boltzmann law. These
findings are in accordance with the ones of the previous example.

\section{Conclusion}\label{sec:conclusion}

In order to describe the time evolution of highly asymmetric systems,
involving widely different length and time scales, like colloidal
dispersions, we have extended the Lattice-Boltzmann formalism for the
description of fluid flow by replacing the standard BGK collision
operator by a discretized Fokker-Planck operator to account for the
dissipative coupling of large solutes to a continuum solvent, without
resolving the molecular scale of the latter. Using an expansion
of the continuous one-particle distribution function in a truncated
Gauss-Hermite basis, as well as standard quadratures with appropriately
chosen weights, we were able to reduce the initial continuous Fokker-Planck
equation to a simple matrix form. The stability of the discrete time
evolution is determined by the diagonal elements of the triangular
collision matrix, which are proportional to the friction coefficient
$\gFP$. A standard Chapman-Enskog expansion leads back to the usual
conservation equations derived from the continuous FP equation in the limit
$\Dt\rightarrow 0$. For finite time steps $\Dt$, the correct continuum
equations are recovered by properly redefining the hydrodynamic variables,
i.e. by introducing the starred current and stress tensor of 
Eqs.~\eref{defjPst}. This leads then to the Lattice Fokker-Planck
algorithm of sec.~\ref{sec:algo}.

This algorithm was first tested against known analytical results for the
time evolution of simple model systems. The numerical efficiency was tested 
in the non-trivial case of colloid sedimentation in the presence of gravity
and a centrifugal force. We intend to use the LFP algorithm to investigate
ion translocation through heterogeneous nanopores, ion transport in 
swollen clays, and various applications in dissipative colloid dynamics.
These applications will benefit from the extensive experience gained over
the years with the related LB method.

Due to the well-known mapping between the Fokker-Planck and the imaginary
time Schr\"{o}dinger equation~\cite{Risken}, the present LFP scheme is also
applicable to the solution of ground-state quantum problems. 

The present LFP scheme has a number of limitations. First of all, the
Fokker-Planck equation itself is never fully rigorous, since a separation of
time scales is never complete, as had already been recognized by
H.A. Lorentz~\cite{BH00} so that non-markovian corrections are always
present~\cite{BP97}. Secondly, to account for collisions between particles, a
BGK term may be added to the discrete FP operator, as stressed several times in
this paper, and illustrated in two of the numerical examples
(cfr. \fig{fig:trans1d} and \ref{fig:trans2d}). However it is not clear how
such a term could account for the long-range hydrodynamic interactions between
Brownian particles induced by the solvent back-flow.  A third limitation emerges
from the stability analysis, which restricts the range of possible values of the
inverse time scales $\gFP$ (associated with the FP operator) and $\tauinv$
(characterizing the BGK operator).  Clearly there is a need for an algorithm
valid to higher order in $\Dt$.  Work to improve along these lines the
Lattice-Fokker-Planck method put forward in this paper is in progress.

\begin{acknowledgements}

D.M. acknowledges financial support from the University of Rome 'La Sapienza'
and from Schlumberger Cambridge Research.  B.R. acknowledges financial support
from the Ecole Normale Sup\'{e}rieure and the Agence Nationale pour la Gestion
des D\'{e}chets Radioactifs (ANDRA, France).

\end{acknowledgements}

\appendix

\section{$\Dim$-Dimensional Hermite Polynomials}\label{sec:hermitepoly}

A complete set of orthonormal polynomials in $\Dim$ variables can be obtained by
products of Hermite polynomials in a single variable.  A detailed presentation
can be found in the excellent work of Grad~\cite{Grad49}. Here we sketch the
basic notions and concentrate on the relations which are useful in the present
work.

Consider the space of real functions $f(\Rd)$ of $\Dim$ variables for which the
integral $\int d\Vd \weD(\Vd) f(\Vd)^2$ exists, where $\weD(\Vd)$ is the
gaussian weight function defined by Eq.~\eref{eq:gaussweight}.  A
$\Dim$-dimensional Hermite polynomial of order $\KiA$ is a tensor
$\HeHomo{\vecK{\DiA}}{\KiA}(\vecd{x})$ of rank $\KiA$. Each component is a
polynomial function in this space. These polynomials form an orthogonal set in
the sense
\beq\label{eq:HeHomoNorm}
\int d\Vd \weD(\Vd) \frac{1}{\vT^{\KiA+\KiB}} 
\HeHomo{\vecK{\DiA}}{\KiA}(\Vd) \HeHomo{\vecK{\DiB}}{\KiB}(\Vd) = 
\delta_{\KiA\KiB}\delta^{(\KiA)}_{\vecK{\DiA}\vecK{\DiB}}
\eeq
where the quantity $\delta^{(\KiA)}_{\vecK{\DiA}\vecK{\DiB}}$ is zero unless the
subscripts $\vecK{\DiA}=\DiA_1\ldots\DiA_\KiA$ are a permutation of
$\vecK{\DiB}=\DiB_1\ldots\DiB_\KiA$. It is a sum of $\KiA!$ terms, each one
being a product of $\KiA$ Kronecker $\delta$'s with the subscripts given by the
all possible permutations of indices from the two sets $\vecK{\DiA}$ and
$\vecK{\DiB}$. The first few polynomials read
\footnote{The Hermite polynomials 
$\HeHomo{\protect\vecK{\DiA}}{\KiA}(\Vd)$ defined here can be considered as a
dimensional-explicit version of the polynomials
$\He{\protect\vecK{\DiA}}{\KiA}(\vecd{\Vd})$ defined by Grad~\cite{Grad49}. The
relation $\HeHomo{\protect\vecK{\DiA}}{\KiA}(\vecd{\Vd}) =
\vT^\KiA \He{\protect\vecK{\DiA}}{\KiA}(\vecd{\Vd}/\vT)$ exists between the two,
where $\vT$ has the dimension of the velocity $\Vd$.}
\bea
\HeHomo{}{0}(\Vd) &=& 1 \\
\HeHomo{\DiA}{1}(\Vd) &=& \Vb_\DiA \\
\HeHomo{\DiA\DiB}{2}(\Vd) &=& \Vb_\DiA \Vb_\DiB-\vT^2\delta_{\DiA\DiB} \\
\HeHomo{\DiA\DiB\DiC}{3}(\Vd) &=& \Vb_\DiA \Vb_\DiB \Vb_\DiC - \vT^2
 (\Vb_\DiA\delta_{\DiB\DiC}+\Vb_\DiB\delta_{\DiA\DiC}
  +\Vb_\DiC\delta_{\DiA\DiB}) \\
\HeHomo{\DiA\DiB\DiC\DiD}{4}(\Vd) &=& \Vb_\DiA \Vb_\DiB \Vb_\DiC \Vb_\DiD 
 - \vT^2 (\Vb_\DiA \Vb_\DiB \delta_{\DiC\DiD} 
 + \Vb_\DiA \Vb_\DiC \delta_{\DiB\DiD} + \\
&&  
  \Vb_\DiA \Vb_\DiD \delta_{\DiB\DiC} + 
  \Vb_\DiB \Vb_\DiC \delta_{\DiA\DiD} + 
  \Vb_\DiB \Vb_\DiD \delta_{\DiA\DiC} + 
  \Vb_\DiC \Vb_\DiD \delta_{\DiA\DiB}) + \nonumber\\
&&
 \vT^4 (\delta_{\DiA\DiB}\delta_{\DiC\DiD}+\delta_{\DiA\DiC}\delta_{\DiB\DiD}+
  \delta_{\DiA\DiD}\delta_{\DiB\DiC}) \nonumber
\eea
Hermite polynomials form a complete set and the expansion \eref{eq:HermiExpa0}
is valid, where the coefficients are given by Eq.~\eref{eq:HermiCoeff}.  In
$\Dim=1$ the polynomials defined here reduce to the so-called Hermite-Chebyshev
polynomials, which differ in normalization from the usual Hermite
polynomials~\cite{AbramowitzStegun}.
The derivative of Hermite polynomials satisfies two important properties.  The
first relates an Hermite polynomial of degree $\KiA$ to one of degree $\KiA-1$
\beq\label{eq:Hprop2}
\partial_{\Vb_\DiB}\HeHomo{\vecK{\DiA}}{\KiA}(\Vd)=
 \delta_{\DiB\DiA_1}\HeHomo{\DiA_2\ldots\DiA_\KiA}{\KiA-1}(\Vd)+
 \ldots+
 \delta_{\DiB\DiA_\KiA}\HeHomo{\DiA_1\ldots\DiA_{\KiA-1}}{\KiA-1}(\Vd)
\eeq
The second is the recurrence relation
\beq\label{eq:Hprop}
\partial_{\Vb_\DiB}\HeHomo{\vecK{\DiA}}{\KiA}(\Vd)=
\frac{1}{\vT^2}[\Vb_\DiB\HeHomo{\vecK{\DiA}}{\KiA}(\Vd)
 -\HeHomo{\DiB\vecK{\DiA}}{\KiA+1}(\Vd)]
\eeq
where $\DiB\vecK{\DiA}$ denotes the $\KiA+1$ indices
$\DiB\DiA_1\ldots\DiA_\KiA$. 

Making use of the quadratures, Eqs.~\eref{eq:quadra} in sec.~\ref{sec:quadra},
integrals of products of Hermite polynomials and a gaussian can be rewritten as
discrete sums on lattice vectors.  The maximum order of the polynomial involved
is dictated by the order of the quadratures. In the practical case of interest
here, a quadrature of order 4 is used in the model of Eq.~\ref{eq:LB2}, where
$\Oherm=2$.  The orthonormality relations Eq.~\eref{eq:HeHomoNorm} become then
the formulae
\bea
\sum_\QiA \wi_\QiA &=& 1 \\
\sum_\QiA \wi_\QiA \Vb_{\QiA\DiA}\Vb_{\QiA\DiB}&=&\vT^2\delta_{\DiA\DiB} \\
\sum_\QiA \wi_\QiA (\Vb_{\QiA\DiA}\Vb_{\QiA\DiB}-\vT^2\delta_{\DiA\DiB}) 
(\Vb_{\QiA\DiC}\Vb_{\QiA\DiD}-\vT^2\delta_{\DiC\DiD}) &=& 
\vT^4(\delta_{\DiA\DiC}\delta_{\DiB\DiD}+
 \delta_{\DiA\DiD}\delta_{\DiB\DiC}) \label{eq:Hermiquadra}
\eea
and the remaining combinations, such as $\sum_\QiA \wi_\QiA \Vb_{\QiA\DiA}$ etc.
are simply 0. From the last Eq.~\eref{eq:Hermiquadra} we can derive the 
the fourth-order tensor formula
\bea
\sum_\QiA \wi_\QiA \Vb_{\QiA\DiA}\Vb_{\QiA\DiB}\Vb_{\QiA\DiC}\Vb_{\QiA\DiD} &=&
\vT^4(\delta_{\DiA\DiB}\delta_{\DiC\DiD}+\delta_{\DiA\DiC}\delta_{\DiB\DiD}+
 \delta_{\DiA\DiD}\delta_{\DiB\DiC}) 
\eea

\section{Eigenfunctions of the $\Dim$-dimensional 
Fokker-Planck operator}\label{sec:eigenhermite}

The $\Dim$-dimensional Hermite polynomials defined in the previous
section can be used to construct eigenfunctions of the
Fokker-Planck operator $\CopB^{FP}[\fB]$ of Eq.~\eref{eq:FPE1} 
in the form of products $\weD(\Vd)\HeHomo{\vecK{\DiA}}{\KiA}(\Vd)$.

Using the fact that for a gaussian
\beq\label{eq:Dgauss}
\partial_{\Vb_\DiA} \weD(\Vd) = -\frac{\Vb_\DiA}{\vT^2}\weD(\Vd)
\eeq
we can write for the action of $\CopB^{FP}$ on these functions 
\bea
\CopB^{FP}[\weD(\Vd)\HeHomo{\vecK{\DiA}}{\KiA}(\Vd)] 
&=& 
 \gFP \partial_{\Vb_\DiB}(\Vb_\DiB + \vT^2\partial_{\Vb_\DiB})  
 [\weD(\Vd)\HeHomo{\vecK{\DiA}}{\KiA}(\Vd)] \nonumber \\
&=&
 \gFP\weD(\Vd)(-\Vb_\DiB\partial_{\Vb_\DiB}+
 \vT^2\partial_{\Vb_\DiB}\partial_{\Vb_\DiB})
 \HeHomo{\vecK{\DiA}}{\KiA}(\Vd)
\eea
Because of relation~\eref{eq:Hprop}, we can write
\beq
\vT^2\partial_{\Vb_\DiB}\partial_{\Vb_\DiB}\HeHomo{\vecK{\DiA}}{\KiA}(\Vd)
=\Dim\HeHomo{\vecK{\DiA}}{\KiA}(\Vd)
 +\Vb_\DiB\partial_{\Vb_\DiB}\HeHomo{\vecK{\DiA}}{\KiA}(\Vd)
 -\partial_{\Vb_\DiB}\HeHomo{\DiB\vecK{\DiA}}{\KiA+1}(\Vd)
\eeq
Using then property~\eref{eq:Hprop2} it is easy to prove that
\beq
\partial_{\Vb_\DiB}\HeHomo{\DiB\vecK{\DiA}}{\KiA+1}(\Vd)=
 -(\KiA+\Dim)\HeHomo{\vecK{\DiA}}{\KiA}(\Vd)
\eeq
Bringing all the above relations together we get
\beq
\CopB^{FP}[\weD(\Vd)\HeHomo{\vecK{\DiA}}{\KiA}(\Vd)] =
 -\gFP\KiA\weD(\Vd)\HeHomo{\vecK{\DiA}}{\KiA}(\Vd)
\eeq
which is the eigenvalue property we wanted to prove. From this 
it is immediate to prove relation~\eref{eq:HcoeffC1} using
\eref{eq:Cherm} and the orthonormality
of the polynomials.

We can use the above results to prove also relation~\eref{eq:HcoeffC2}.
From \eref{eq:Dgauss} and \eref{eq:Hprop} we get
\bea
\CopB^{EXT}[\weD(\Vd)\HeHomo{\vecK{\DiA}}{\KiA}(\Vd)] 
&=& -\aEF_\DiB\partial_{\Vb_\DiB}
 [\weD(\Vd)\HeHomo{\vecK{\DiA}}{\KiA}(\Vd)] \nonumber \\
&=& \frac{\weD(\Vd)}{\vT^2}\aEF_\DiB\HeHomo{\DiB\vecK{\DiA}}{\KiA+1}(\Vd)
\eea
from which, using the expansion~\eref{eq:HermiExpa0},
\bea
\int d\Vd \HeHomo{\vecK{\DiC}}{\KiB}(\Vd) \CopB^{EXT}[\fB] 
&=& \sum_{\KiA=0}^{\infty} \frac{\CHef{\vecK{\DiA}}{\KiA}}{\vT^{2\KiA}\KiA!}
 \int d\Vd \HeHomo{\vecK{\DiC}}{\KiB}(\Vd) 
 \CopB^{EXT}[\weD(\Vd) \HeHomo{\vecK{\DiA}}{\KiA}(\Vd)] \nonumber \\
&=& \sum_{\KiA=0}^{\infty} \frac{\CHef{\vecK{\DiA}}{\KiA}}{\vT^{2\KiA}\KiA!}
 \frac{\aEF_\DiB}{\vT^2} 
 \int d\Vd \HeHomo{\vecK{\DiC}}{\KiB}(\Vd) 
 \weD(\Vd)\HeHomo{\DiB\vecK{\DiA}}{\KiA+1}(\Vd)
 \nonumber \\
&=& \frac{1}{(\KiB-1)!} \CHef{\vecK{\DiA}}{\KiB-1} 
 \aEF_\DiB \delta_{\vecK{\DiC},\DiB\vecK{\DiA}}^{(\KiB)}
 \nonumber \\
&=& \aEF_{\DiC_1}\CHef{\DiC_2\ldots\DiC_\KiB}{\KiB-1} + \ldots +
 \aEF_{\DiC_\KiB}\CHef{\DiC_1\ldots\DiC_{\KiB-1}}{\KiB-1}
\eea
where we used Eq.~\eref{eq:HeHomoNorm} and 
the fact that $\CHef{}{\KiB-1}$ is invariant under permutations
of its $\KiB-1$ indices.



\newpage

\newpage

\newpage

\newpage

\newpage

\newpage

\end{document}